\documentclass[usenatbib]{mn2e}
\usepackage{graphicx}
\usepackage{tablefootnote}

\usepackage{natbib}
\graphicspath{{figures/}{./}}

\begin{document}

\newcommand{\apj}{ApJ}
\newcommand{\apjl}{ApJ}
\newcommand{\apjs}{ApJS}
\newcommand{\aap}{A\&A}
\newcommand{\aj}{AJ}
\newcommand{\mnras}{MNRAS}
\newcommand{\pasp}{PASP}
\newcommand{\pasj}{PASJ}
\newcommand{\physrep}{Physics Reports}
\newcommand{\prd}{PRD}
\newcommand{\prl}{PRL}
\newcommand{\nat}{Nature}
\newcommand{\araa}{ARAA}

\def\eps@scaling{1.0}%
\newcommand\epsscale[1]{\gdef\eps@scaling{#1}}%
\newcommand\plotone[1]{%
 \centering
 \leavevmode
 \includegraphics[width={\eps@scaling\columnwidth}]{#1}%
}%
\newcommand\plottwo[2]{%
 \centering
 \leavevmode
 \columnwidth=1.0\columnwidth
 \includegraphics[width={\eps@scaling\columnwidth}]{#1}%
 \hfil
 \includegraphics[width={\eps@scaling\columnwidth}]{#2}%
}%

\newcommand\acknowledgments[1]{#1}
\newcommand\facility[1]{#1}

\newcommand\ion[2]{#1$\;${\small{#2}}}

\newcommand{\cooname}{J141446.82+544631.9}
\newcommand{\fullname}{CFHTLS \cooname}
\newcommand{\lae}{J1414+5446}
\newcommand{\Msun}{M_{\sun}}
\newcommand{\Msunyr}{M_{\sun}~{\rm yr}^{-1}}
\newcommand{\flam}{{\rm erg}~{\rm s}^{-1}~{\rm cm}^{-2}~{\rm \AA}^{-1}}
\newcommand{\fcgs}{{\rm erg}~{\rm s}^{-1}~{\rm cm}^{-2}}
\newcommand{\ergss}{{\rm erg}~{\rm s}^{-1}}
\newcommand{\kms}{{\rm km}~{\rm s}^{-1}}
\newcommand{\laez}{5.424}
\newcommand{\slsid}{SL2S J141447+544703}
\newcommand{\astrodrizzle}{\textsc{AstroDrizzle}}
\newcommand{\gdssrc}{GDS J033218.92-275302.7}

\title[A bright $z=5.4$ galaxy]{A bright lensed galaxy at $z=5.4$ with strong Ly$\alpha$ emission}
\author[McGreer et al.]{
Ian D. McGreer,$^{1,2,}$\thanks{Email: imcgreer@as.arizona.edu} 
Benjamin Cl\'ement,$^{3}$ 
Ramesh Mainali,$^{1}$
Daniel P. Stark,$^{1}$ \newauthor
Max Gronke,$^{4}$ 
Mark Dijkstra,$^{4}$
Xiaohui Fan,$^{1}$
Fuyan Bian,$^{5}$
Brenda Frye,$^{1}$ \newauthor
Linhua Jiang,$^{6}$
Jean-Paul Kneib,$^{7,8}$ 
Marceau Limousin,$^{8}$ 
Gregory Walth $^{9}$\\
$^{1}$ Steward Observatory, The University of Arizona, 
                 933 North Cherry Avenue, Tucson, AZ 85721--0065 \\
$^{2}$ Spaceflight Industries, 1505 Westlake Ave N Suite 600, Seattle, WA 98109 \\
$^{3}$ Univ Lyon, Univ Lyon1, Ens de Lyon, CNRS, Centre de Recherche Astrophysique de Lyon UMR5574, F-69230, Saint-Genis-Laval, France \\
$^{4}$ Institute of Theoretical Astrophysics, University of Oslo, Postboks 1029, 0315 Oslo, Norway \\
$^{5}$ Research School of Astronomy \& Astrophysics, Australian National University, Canberra, ACT, 2611 Australia \\
$^{6}$ Kavli Institute for Astronomy and Astrophysics, Peking University, Beijing 100871, China \\
$^{7}$ Institute of Physics, Laboratory of Astrophysics, Ecole Polytechnique F\'ed\'erale de Lausanne (EPFL), Observatoire de Sauverny, 1290 Versoix, Switzerland \\
$^{8}$ Aix Marseille Univ, CNRS, CNES, LAM, Laboratoire d'Astrophysique de Marseille, Marseille, France \\
$^{9}$ University of California, Center for Astrophysics and Space Sciences, 9500 Gilman Drive, San Diego, CA 92093, USA
}

\maketitle

\begin{abstract}
We present a detailed study of a unusually bright, lensed galaxy at $z=\laez$ 
discovered within the CFHTLS imaging survey. With an observed flux of 
$i_{\rm AB}=23.0$, \cooname\ is one of the brightest galaxies known at 
$z>5$. It is characterized by strong Ly$\alpha$ emission, reaching a peak
in (observed) flux density of $>10^{-16}~\flam$. A deep 
optical spectrum from the LBT places strong constraints on \ion{N}{V} and 
\ion{C}{IV} emission, disfavouring an AGN source for the emission. However, 
a detection of the \ion{N}{IV}]~$\lambda$1486 emission line indicates a hard 
ionizing continuum, possibly from hot, massive stars.  Resolved imaging from 
{\sl HST} deblends the galaxy from a foreground interloper; these observations 
include narrowband imaging of the Ly$\alpha$ emission, which is marginally 
resolved on $\sim$few~kpc scales and has EW$_0~\sim$ 260\AA. The Ly$\alpha$ 
emission extends over $\sim2000~\kms$ and is broadly consistent with expanding 
shell models. SED fitting that includes {\sl Spitzer}/IRAC photometry suggests 
a complex star formation history that include both  a recent burst and an 
evolved population. \lae\ lies 30\arcsec\ from the centre of a known lensing 
cluster in the CFHTLS; combined with the foreground contribution this leads to 
a highly uncertain estimate for the lensing magnification in the range 
$5 \la \mu \la 25$. Because of its unusual brightness \lae\ affords unique 
opportunities for detailed study of an individual galaxy near the epoch of 
reionization and a preview of what can be expected from upcoming wide-area 
surveys that will yield hundreds of similar objects.
\end{abstract}


\section{Introduction}\label{sec:intro}

The census of star-forming galaxies near the reionization epoch has expanded
greatly in recent years, primarily due to deep imaging surveys at optical
and near-infrared wavelengths that capture the rest-frame ultraviolet
emission produced by ongoing star formation \cite[see review by][]{Stark16}.
These surveys generally fall into two classes.
Lyman Break Galaxies (LBGs) are colour-selected based on the sharp break
in the continuum flux at the Ly$\alpha$ wavelength induced by the absorption
of intervening neutral hydrogen at $z\ga4$.
A subset of high-redshift galaxies are characterized by
strong Ly$\alpha$ emission and are classified as Lyman-alpha Emitters (LAEs);
they have traditionally been identified by the flux excess present in a narrow bandpass
designed to capture the line emission, although powerful IFUs such as MUSE are 
increasingly being used to conduct line surveys \citep[e.g.,][]{Smit+17}.
Deep surveys with the Hubble Space Telescope (HST) 
have been critical in building large samples of faint galaxies at $z>5$, now numbering 
in the thousands \citep{Bouwens+14}; although extrapolation beyond the 
observed population is required for galaxies to provide sufficient
ionizing photons to completely reionize 
the diffuse intergalactic gas \citep[e.g.,][]{Robertson+15}.

Broadly speaking, both LBG and LAE surveys at high redshift probe  
relatively small volumes. 
The most recent ultra-deep survey field from HST has an area of $<5$~arcmin$^2$ 
\citep[UDF12:][]{Ellis+13,Koekemoer+13}; in total the
HST surveys have an area $< 0.3$~deg$^2$ \citep{Bouwens+14}.
Recent ground-based surveys have yielded rarer, brighter LBGs in areas 
covering a few sq. deg. \citep{Willott+13,Bowler+15}. LAE surveys can 
be successfully conducted from the ground as the narrow filter can be 
placed in regions of relatively low sky background between prominent OH 
sky emission bands \citep[e.g.,][]{Kashikawa+04}; the Hyper Suprime-Cam
(HSC) surveys with the Subaru telescope are now probing LAEs at 
$z=6~\mbox{--}7$ over areas of tens of square degrees to exquisite depth
\citep{Ouchi+17}. However, by employing a narrow bandpass these surveys are 
restricted to thin redshift slices and hence relatively small volumes 
($\sim0.5$\ Gpc$^3$ for the HSC surveys). 
Blind spectroscopic surveys with slits or IFUs cover a 
much wider redshift range, but are limited to small areas.

Because of the limited volume, galaxies discovered in the
aformentioned surveys tend to be faint and difficult to study in detail.
Bright galaxies can be examined with spectroscopic and multiwavelength 
observations, probing the physical conditions in individual systems 
\citep[e.g.,][]{Bayliss+14,Yang+14,Smit+17}
with tools otherwise limited to stacking analyses of large numbers of
photometric galaxies \citep[e.g.,][]{JSE12}. The profiles of both the 
Ly$\alpha$ emission line \citep[e.g.,][]{GBD15} and interstellar absorption 
lines \citep[e.g.,][]{Shapley+03} probe the covering fraction and kinematics 
of neutral gas, including large-scale outflows. Metal lines provide 
constraints on the ionizing radiation field in early galaxies, which may be 
driven by hard radiation from very hot, metal-poor massive star populations
\citep{Mainali+17,Stark+15ciii,Stark+15civ}. Finally, mid-infrared photometry 
with the {\it Spitzer Space Telescope} constrains the stellar mass of bright 
galaxies and the presence of an evolved population of stars, indicating 
previous bursts of star formation.

We have discovered an extremely bright galaxy at $z=5.426$ that was
initially selected as a high-redshift quasar candidate. This object
was drawn from a relatively wide-area imaging survey: the CFHTLS-W3 field 
\citep{Gwyn12} covers 49~deg$^2$ to a depth of $i=25.7$. 
With $i_{\rm AB}=23.0$, it is (to the best of our knowledge) 
the brightest galaxy known at $z>5$, with only a handful of galaxies
even comparable in observed flux
(e.g., the lensed galaxy A1689\_{2} at $z=4.87$ has 
${\rm I}_{\rm AB}=23.3$ [\citealt{Frye+02}],
the LAE J0335 at $z\sim5.7$ has ${\rm I}_{\rm AB}=24.3$ [\citealt{Yang+14}]
).
The galaxy, \fullname\ (hereafter \lae), is characterized by a strong
Ly$\alpha$ emission line and a rare detection of \ion{N}{IV}~$\lambda$1468 
emission. Whether it is \emph{intrinsically} luminous is not clear,
as its magnification due to gravitational lensing is poorly constrained.

The structure of the paper is as follows. First, in \S\ref{sec:obs} we 
present an array of multiwavelength observations of the galaxy, including
optical spectroscopy that confirms its redshift and characterizes the
strong Ly$\alpha$ emission (\S\ref{sec:obs_mmt}~and~\S\ref{sec:obs_lbt}); 
infrared photometry that probes the rest-frame optical emission
(\S\ref{sec:obs_spitzer}); and both new and archival HST imaging that 
resolves the emission into multiple components (\S\ref{sec:obs_hst}). 
Multi-band image decomposition is used to characterize the observed 
morphology as detailed in \S\ref{sec:imagemodel}. In \S\ref{sec:discuss} we 
explore the physical properties of this galaxy: the spatial extent of the UV 
continuum and Ly$\alpha$ emission; rest-frame UV spectral properties including 
detailed modeling of the Ly$\alpha$ line profile; estimates of the 
gravitational lensing amplification from simulations of the foreground mass; 
stellar population models from spectral energy distribution (SED) fitting; 
and finally inferences about the star formation and ionizing spectrum. Brief 
conclusions are given in \S\ref{sec:conclude}. All magnitudes are quoted on 
the AB system \citep{OG83} and when needed a flat $\Lambda$CDM cosmology
is adopted with parameters derived from the Planck 2013 results 
\citep[$\Omega_\Lambda=0.692$~and~$H_0=67.8~{\rm km}~{\rm s}^{-1}~{\rm Mpc}^{-1}$,][]{Planck2013}.

\section{Observations}\label{sec:obs}

\subsection{Initial Selection from CFHTLS-W3}\label{sec:selection}

\lae\ was initally selected as a quasar candidate during our survey of faint 
$z\sim5$ quasars in the CFHTLS imaging fields \citep{McGreer+17}. 
We started with the publicly available images and catalogs from the CFHTLS W3 
field \citep{Gwyn12}, providing $ugriz$ photometry over the 49~deg$^2$ field.
We adopted the 2\arcsec\ aperture photometry measurements from the catalogs, 
then selected high redshift quasar candidates using the 
Likelihood method outlined in \citet{Kirkpatrick+11} which assigns quasar
probabilities based on the observed fluxes. 
Further details of our selection method can be found in \citet{McGreer+17}.

\begin{figure*}
\centering
\epsscale{2.0}
\plotone{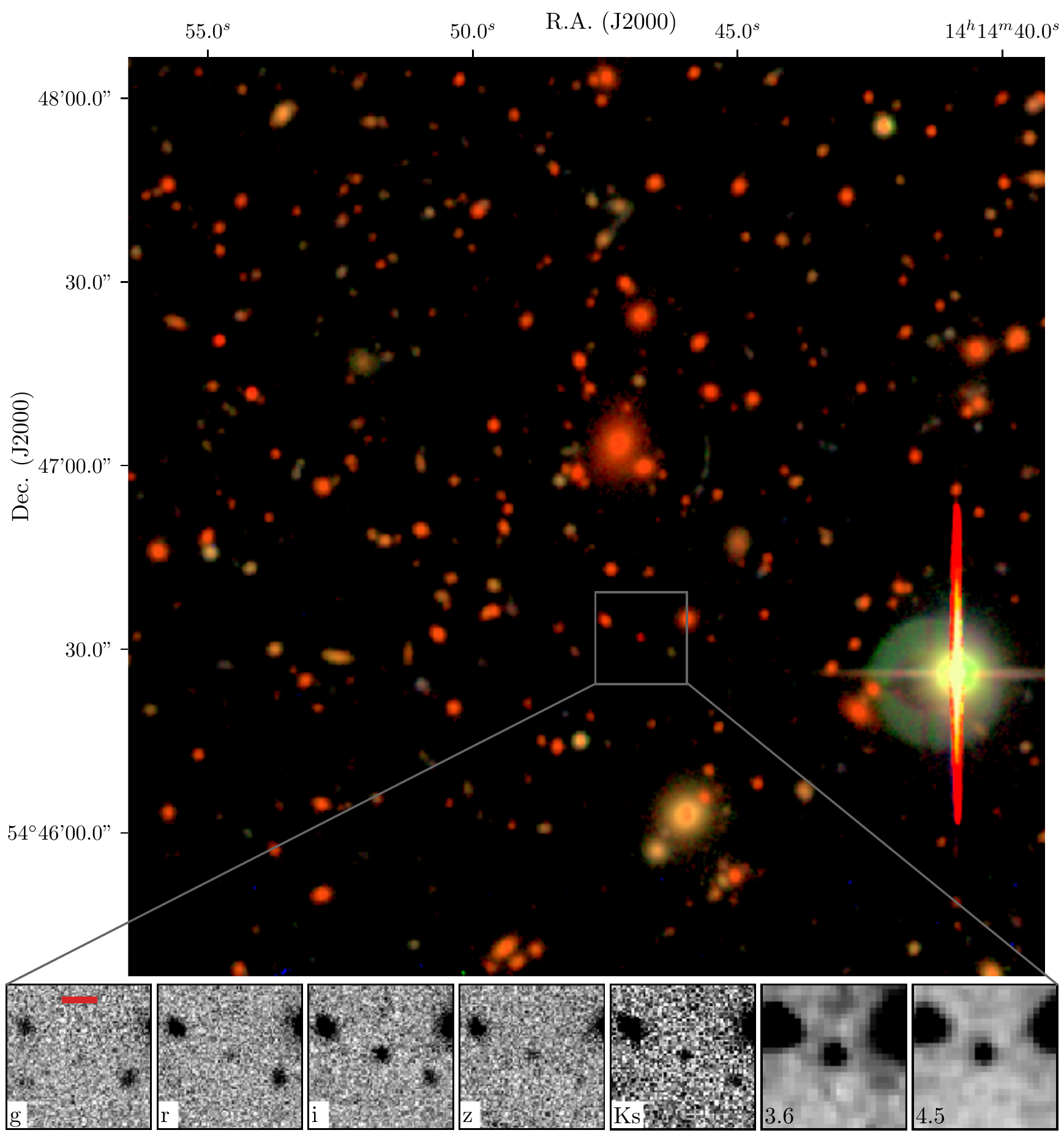}
\caption{Colour image of the $\sim2$\arcmin\ field surrounding \lae. The colour
 image is generated from the CFHT $gri$ images using the \citet{Lupton+04}
 method. The brightest galaxy in the foreground group \slsid\ 
 \citep{Cabanac+07} is clearly visible just above the image centre; as well 
 as the large number of group member galaxies with similar colours. \lae\ lies 
 just below the image centre, highlighted by the box. The panels in the 
 bottom row display cutout images with a size of 15\arcsec\ from the 
 CFHT $griz$, LBT/LUCI $Ks$, and Spitzer/IRAC 3.6$\mu$m and 
 4.5$\mu$m data (left to right). A 3\arcsec\ scalebar is provided in the
 leftmost panel ($g$-band) as a thick red line. \lae\ is not clearly 
 resolved in any image, including the high-$S/N$ $i$-band image which contains 
 the Ly$\alpha$ emission.
 \label{fig:colorim}
 }
\end{figure*}

\lae\ could also have been selected by simple colour criteria; for example, the
colour cuts we adopted in \citet{McGreer+13} to select $z=5$ quasars. 
Although \lae\  proved to be a galaxy, we did not reject it as such using 
morphological criteria from the CFHT imaging. At this flux level 
star/galaxy separation is challenging, at the time \lae\ was targeted we 
mainly eliminated galaxies by visual inspection of the images. Inspection 
of the CFHT images of \lae\ did not indicate that it was resolved; see 
Figure~\ref{fig:colorim}.

\lae\ is covered by two independent pointings in the CFHTLS-W3 field 
(W3+0+0 and W3-0+0). During selection we used the aperture photometry from 
a single field. We have since updated the photometry, first by using 
PSF-shaped fluxes from PSFEx, and second by coadding the two sets of 
measurements. The improved CFHT photometry is provided in 
Table~\ref{tab:phot}. The weak detections in the $u$ and $g$ bands are
unexpected for a galaxy at this redshift. At the time of target selection the 
significance of these detections was inconclusive and thus the object was not
rejected as a high-redshift candidate. This issue will be 
discussed further in \S\ref{sec:foreground}.

\subsection{MMT Observations}\label{sec:obs_mmt}

We first obtained a spectrum of \lae\ on UT 2012 May 28 using the Red Channel
spectrograph on the MMT 6.5m telescope. The object was placed in a 
$1\arcsec~\times~180\arcsec$\ longslit after a blind offset from a nearby
reference star and dispersed with a 270~mm$^{-1}$ grating centred at 7500\AA, 
providing coverage from 5700~\AA\ to 9100~\AA\ at a resolution of $R\sim640$. 
Two 20 min. exposures in good conditions with 0.9\arcsec\ seeing were obtained. 

The spectrum was reduced using standard IRAF tasks. Wavelength calibration 
was obtained from an HeNeAr lamp and then refined using night sky lines in 
the science spectra. Flux calibration was obtained from an observation of
PG1708+602 taken shortly after the science observations.

We obtained a higher resolution spectrum of \lae\ on 2013 May 18 using the
1200-9000 grating on Red Channel. The spectrum extends from 7240~\AA\ to
8050~\AA\ at a dispersion of 0.8~\AA~pix$^{-1}$. We measured a resolution of 
$R\sim3000$ at $\sim7800$\AA\ from unblended night sky lines. We obtained 
two exposures of 1800s each in $\sim1.2$\arcsec\ seeing at position angle 
of 157.6$^\circ$. The spectra were processed with the same routines as for 
the low dispersion spectrum, using observations of standard HZ44 for flux 
calibration. The flux calibration is highly uncertain given the variable 
conditions during the observing period and the likelihood of slight 
mis-centreing of the object in the slit due to the use of a blind offset 
for acquisition.

\subsection{LBT Observations}\label{sec:obs_lbt}

In this section we describe imaging and spectroscopy of \lae\ obtained with
the 2$\times$8.4-m Large Binocular Telescope (LBT). This includes optical
imaging and medium resolution spectroscopy with the MODS1 instrument
\citep{Pogge+06}, and near-infrared imaging with the LUCI1 instrument
\citep{Seifert+03}.

\begin{figure}
\centering
\epsscale{1.0}
\plotone{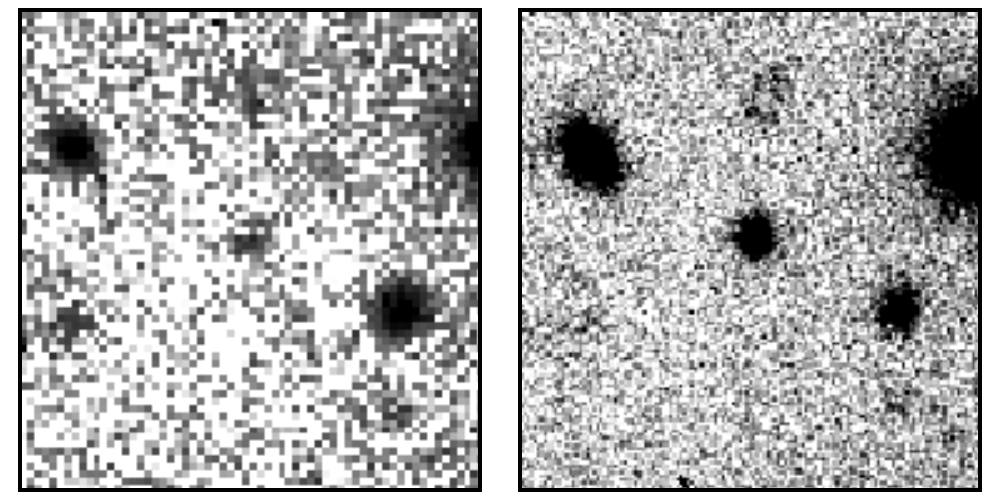}
\caption{LBT/MODS1 $g$ and $i$ band image cutouts of \lae, with the same
 orientation as in Figure~\ref{fig:colorim}.
 The $g$-band image has been rebinned by a factor of two and
 is displayed on an arcsinh scale to enhance the weak detection of \lae.
 Although faint, this clearly confirms the $\sim2\sigma$ flux measured from
 the CFHT image.
 The $i$-band image is displayed at native resolution; \lae\ is detected at
 high $S/N$ and is unresolved in $0\farcs{6}$ seeing.
 \label{fig:modsim}
 }
\end{figure}

\subsubsection{LUCI1 imaging}

We obtained $J$ band imaging with LUCI1 on 2013 Mar 5 in poor, non-photometric 
conditions with variable cloud extinction and $\sim1.4$\arcsec\ seeing. A 
total of 60 min. of integration was accumulated through dithered 6$\times$20s 
individual exposures. Further LUCI1 observations were obtained on 2015 Apr 6-7
using the $K_{\rm s}$ bandpass. The night of 2015 Apr 6 was non-photometric
with passing clouds with 0.7\arcsec seeing. The total exposure
time was 36 min. after rejecting a few integrations that were strongly
affected by low transparency. The following night was photometric and the 
total exposure time was 44 min. with seeing of 0.8\arcsec.

The LUCI1 data were processed in a standard fashion using IRAF tasks, 
incorporating dark current subtraction, flat fields generated from 
combining the science images, and running sky subtraction. The processed 
images for each night were then shifted and combined to construct the 
final images. Individual images were weighted based on the transparency 
($T$), seeing (FWHM), and sky background ($B$) as $T/({\rm FWHM}^2 + B)$ 
when combining, where each weight term is relative to the maximum value of
the given parameter. Finally, the combined images were binned by a factor of 
two along each axis to a resulting pixel scale of 0.24\arcsec. Object 
detection was performed on the binned images using \texttt{SExtractor} 
\citep{sextractor}, and astrometric solutions were obtained by matching 
well-detected objects to the CFHTLS-W3 $i$-band catalog.

\begin{table}
 \begin{center}
 \caption{Photometry of \lae\ based on total fluxes measured through elliptical apertures (\texttt{SExtractor} MAG\_AUTO).}
 \label{tab:phot}
 \begin{tabular}{llc}
 \hline
  Source & Band & AB mag \\
 \hline
      CFHT  &            $u$ & $ 26.45 \pm 0.28 $ \\
            &            $g$ & $ 26.58 \pm 0.23 $ \\
            &            $r$ & $ 25.03 \pm 0.08 $ \\
            &            $i$ & $ 23.00 \pm 0.02 $ \\
            &            $z$ & $ 23.45 \pm 0.09 $ \\
  LBT/MODS  &            $g$ & $ 26.61 \pm 0.30 $ \\
            &            $i$ & $ 22.89 \pm 0.02 $ \\
  LBT/LUCI  &            $J$ & $ 23.77 \pm 0.37 $ \\
            &           $Ks$ & $ 23.14 \pm 0.25 $ \\
    Spitzer &      3.6$\mu$m & $ 22.37 \pm 0.09 $ \\
            &      4.5$\mu$m & $ 22.03 \pm 0.06 $ \\
\hline
 \end{tabular}
 \end{center}
 \medskip
 Note -- Magnitudes are on the AB system \citep{OG83} and have been corrected 
  for extinction using the \citet{SFD98} maps. No attempt has been made to
  deblend the foreground and background galaxies.
\end{table}

We used 2MASS \citep{twomass} stars detected within the field to determine 
the image zero points. The calibration accuracy is severely limited by the 
small number of 2MASS stars available -- only 5 (2) in the $J$ ($Ks$) band. 
We checked the LUCI photometry against red sequence galaxies selected from 
the images, most of which lie at the foreground cluster redshift. Comparing 
our colours to a template red galaxy spectrum at this redshift, we find shifts 
of $\sim0.25$ and $\sim 0.13$ mag in the $J$ and $Ks$ bands, respectively. We 
apply these shifts and add an equal amount of error in quadrature to the 
photometry in order to capture the calibration uncertainty.

\begin{figure*}
\epsscale{2.0}
\plotone{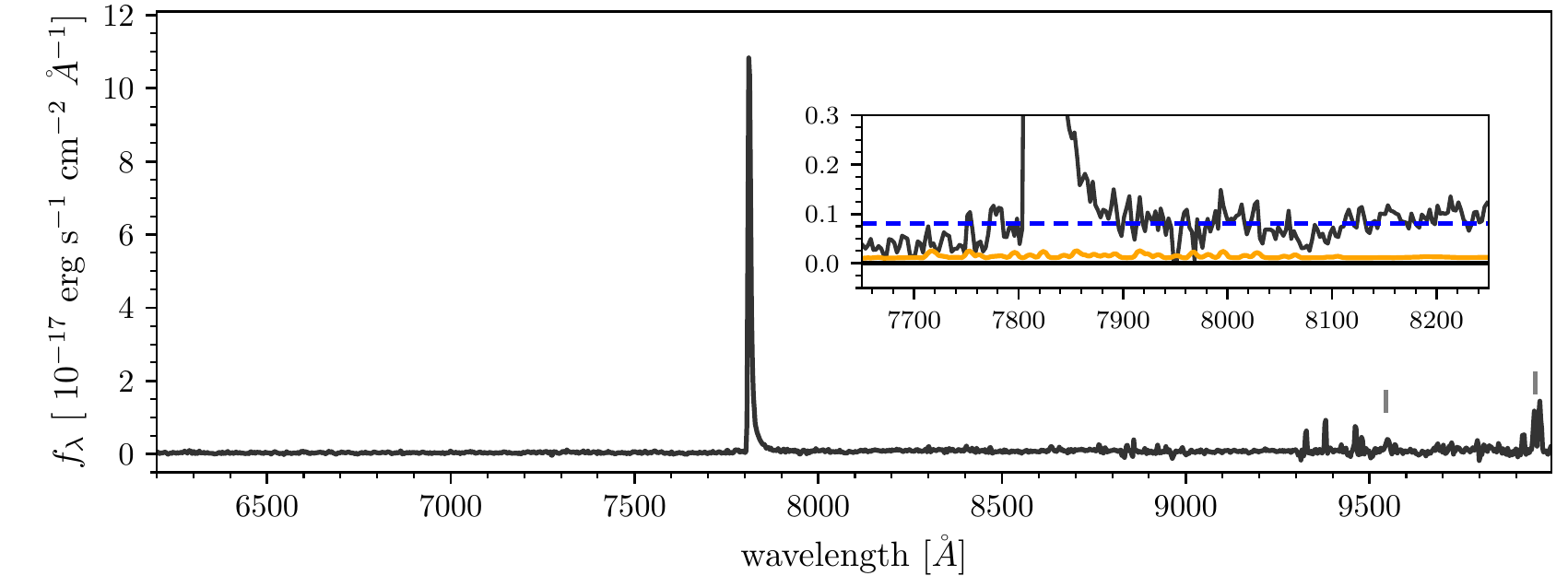}
\caption{LBT/MODS1 spectrum of \lae, dominated by the strong Ly$\alpha$ 
 emission feature. The inset panel highlights the continuum emission 
 detected at $\sim2\sigma~{\rm pix}^{-1}$ redward of the Ly$\alpha$ 
 emission (the continuum fit is indicated by the dashed blue line and the
 error spectrum by the solid orange line). Flux is also detected at 
 $\sim1\sigma~{\rm pix}^{-1}$ blueward of Ly$\alpha$, although there is a 
 clear spectral break at the wavelength Ly$\alpha$. The wavelengths of the
 \ion{N}{IV} and \ion{C}{IV} features are marked with small vertical lines and 
 will be discussed in \S\ref{sec:uvspec}.
 \label{fig:modsspec}
 }
\end{figure*}

\subsubsection{MODS1 imaging}

We observed \lae\ with the imaging mode of MODS1 on 2014 May 31. MODS1 
includes a dichroic for observing blue and red wavelengths simultaneously. 
The field-of-view  in the 3K$\times$3K imaging mode is 6\arcmin\ across, and 
the pixel scale for the blue (red) channel is $0\farcs{120}$ ($0\farcs{123}$). 
We used the $g$ filter in the blue channel and the $i$ filter in the red 
channel, and obtained six dithered exposures with individual integration times 
of 5 min. Conditions were clear and photometric, and the seeing measured from 
the images is $0\farcs{6}$ in the $i$ band and $0\farcs{75}$ in the $g$ band.

Standard methods were employed to process the optical images, using custom
Python routines. A series of bias images were median-combined to create a 
master bias. Pixel flat fields were generated from a series of exposures 
taken with an internal lamp. The final flat field correction consisted
of a stack of the pixel flats, as well as an illumination correction derived
from twilight sky flat images. After applying the bias and flat field
corrections, the individual science images were processed and combined using
\texttt{Scamp} \citep{scamp} to obtain initial astrometric solutions and 
\texttt{SWarp} \citep{Bertin+02} to coadd the images. Sky subtraction was enabled 
when combining with \texttt{SWarp} as the sky level varied substantially over the 
course of the observations. The images were combined in two iterations, the 
first iteration produced a reference image from which cosmic rays and other 
defects in individual images were identified and masked prior to the final 
coaddition.

We produced two sets of final images in each band. The first are in the 
native pixel scale and footprint of each detector. The second are matched 
to the $g$-band pixel scale and aligned using \texttt{SWarp}. We derive object 
catalogs from all images using \texttt{SExtractor}; for the pair of aligned images 
we used dual-image mode with the $i$-band image for detection. Finally, 
we registered the images to the CFHTLS astrometry and determined the
photometric zero point by matching to stars in the CFHTLS catalogs. The 
MODS $g$ and $i$ filters are slightly bluer than the corresponding CFHT 
filters; we thus corrected for a slight tilt ($\la1$\% over the range of 
interest) between the two photometric systems in order to place the LBT 
magnitudes on the CFHT system. The calibrated photometry from the dual-image 
mode catalogs is listed in Table~\ref{tab:phot}. The primary result from this
imaging is confirmation of the $g$-band detection in the CFHTLS.

\subsubsection{MODS1 spectroscopy}\label{sec:lbtmods}

We obtained optical spectroscopy of \lae\ with MODS1 on three different nights.
The goals of the spectroscopy were twofold. First, to obtain a deep spectrum
of the high-$z$ galaxy to search for weak emission and absorption features.
Second, we targeted the lensed arc candidates as well as cluster member 
galaxies in order to better constrain the lensing model. We designed three 
slitmasks, each of which included the \lae\ as a target. Two of the masks
included slits for a long tangential arc \citep[T1 in][]{Cabanac+07} and one mask 
included both of the candidate radial arcs \citep[R1 and R2 in][]{More+12}.
Details of the MODS1 observations are provided in Table~\ref{tab:modsobs}\ and
results from the multi-object spectroscopy are described in 
Appendix~\ref{sec:appendix}.

\begin{table}
 \begin{center}
 \caption{LBT/MODS1 multi-slit spectroscopic observations. The final column
 notes which of the lensed arc candidates described in \S\ref{sec:lbtmods}
 are included on each mask.}
 \label{tab:modsobs}
 \begin{tabular}{cccll}
 \hline
  Mask ID & UT & Exp. (hr) & conditions & notes \\
 \hline
 505919 & 2013-04-13 & 1.9 & 0.8\arcsec, passing clouds & T1 \\
 523405 & 2015-03-25 & 3.8 & 1.0\arcsec, clear & T1 \\
 510122 & 2015-03-26 & 3.7 & 0.6--1.0\arcsec, cloudy & R1, R2 \\
\hline
 \end{tabular}
 \end{center}
\end{table}

All observations employed the dual grating mode of MODS1 for complete wavelength 
coverage from 3200~\AA\ to 1~$\mu$m. The slits for \lae\ were 14--20\arcsec\ in
length while the slits for the galaxy targets were 7--10\arcsec; all slits had
a width of 1\arcsec. The resolution in the blue channel (up to 6000~\AA) is 
$R\sim1100$ and in the red channel (starting at 5000~\AA) is $\sim1400$.

The spectra were processed with Version 2.0 of the \texttt{modsCCDRed}
software, and Version 0.2p1 of the \texttt{modsIDL} spectral
reduction pipeline\footnote{http://www.astronomy.ohio-state.edu/MODS/Software/modsIDL/}.
Individual frames were bias-subtracted and then flat-fielded using a series 
of internal lamp calibrations. Wavelength calibration was provided by a 
combination of internal arcs and has an rms $<0.1$\AA.
Flux calibration was obtained through observations of the spectrophotometric
standard stars Feige 34 and BD+33d2642 acquired in the same night. The individual 
science frames were sky-subtracted and one-dimensional spectra were extracted 
using \texttt{modsIDL}. The \texttt{modsIDL} implementation uses boxcar extraction; 
we constructed model profiles using stars included in the slit masks 
and used these models to perform optimal extraction \citep{Horne86}, with a 
significant gain in $S/N$ compared to the boxcar-extracted spectra.
The individual spectra from each night were combined with inverse-variance 
weighting, scaling the Ly$\alpha$ flux measured from each image to account
for transparency variations. Finally, the spectra from the three nights were 
combined to produce the final MODS spectrum shown in Figure~\ref{fig:modsspec}. 
Before combining, a correction to the flux calibration for the spectra from each
mask was determined by comparing the spectro-photometry obtained for the field
galaxies to their photometry in the CFHTLS. The total exposure time is 
$\sim$9.4 hours; however, two of the nights were affected by passing clouds and 
we estimate the effective exposure time to be $\sim7$ hours.

\subsection{HST Imaging}\label{sec:obs_hst}

\subsubsection{Archival WFPC2 observations}\label{sec:hst_wfpc2}

\lae\ is within the field of \slsid, a lensing group included in the 
Strong Lensing Legacy Survey \citep[SL2S;][]{Cabanac+07}. This survey 
identified candidate strong lenses up to $z\sim1$ in the CFHTLS imaging, 
and included HST observations as part of a Cycle 16 SNAPSHOT program (GO 
\#11289, PI: Kneib). The \slsid\ field was observed with three dithered 
400s exposures using the WFPC2 instrument and the F606W bandpass. \lae\ is 
located $\sim30$\arcsec\ from the centre of the lensing group and was well 
within the WFCP2 image.
The F606W bandpass is entirely blueward of Ly$\alpha$ at $z=5.4$ and thus 
samples the attenuated rest-FUV continuum of the galaxy.

We processed the archival WFPC2 images using 
\textsc{AstroDrizzle} \footnote{http://drizzlepac.stsci.edu/}, 
setting pixfrac=1.0 and drizzling to a final
pixel scale of $0{\farcs}{05}$. We found the default astrometry matched 
the CFHT astrometry to an accuracy $<0{\farcs}{05}$ . 
\lae\ is clearly detected in the final WFPC2 mosaic; furthermore, on 
$<0{\farcs}4$ scales it resolves into up to four components 
(Figure~\ref{fig:hstims}).

\subsubsection{Cycle 22 ACS/WFC3 observations}\label{sec:hst_cycle22}

We obtained Cycle 22 imaging of \lae\ 
on 2015 Nov 14 with ACS and WFC3 (GO \#13762, PI: McGreer). We employed 
three bandpasses to probe the rest-UV continuum from the high-$z$ galaxy: 
ACS/F850LP and WFC3-IR F125W and F160W, spanning $\sim1400\mbox{-}2600$\AA\ 
rest-frame. We also used the ACS ramp filter FR782N centred at 7816\AA\ to 
capture the Ly$\alpha$ emission. The total exposure times were 813s for the
ramp filter, 1011s for ACS/F850LP, and 1359s each for the two WFC3
bands. All observations used a standard three-point dither pattern.
We imposed an orientation constraint on the ACS observations to ensure 
that the F850LP image would include the brightest cluster galaxy and the 
lensed arcs.

Images were processed using \astrodrizzle. A first-pass drizzled ACS/F850LP 
image was registered to the CFHT $i$-band astrometry as with the WFPC2 image. 
The F850LP and FR782N images were obtained in the same orbit with the same 
orientation and field centre, thus the updated astrometry was propagated back 
into all the ACS images using the \textsc{tweakshifts} routine before 
generating a final mosaic. We used pixfrac=0.8 and a scale of 0\farcs{03} 
for the output pixel grid \citep{Koekemoer+11}. The WFC3 images were processed
in a similar fashion, but with a final scale of 0\farcs{06}. Comparing
object centroids between the various HST bands we find they are aligned to
$\la$0\farcs{05} accuracy.

\begin{figure*}
\centering
\epsscale{2.0}
\plotone{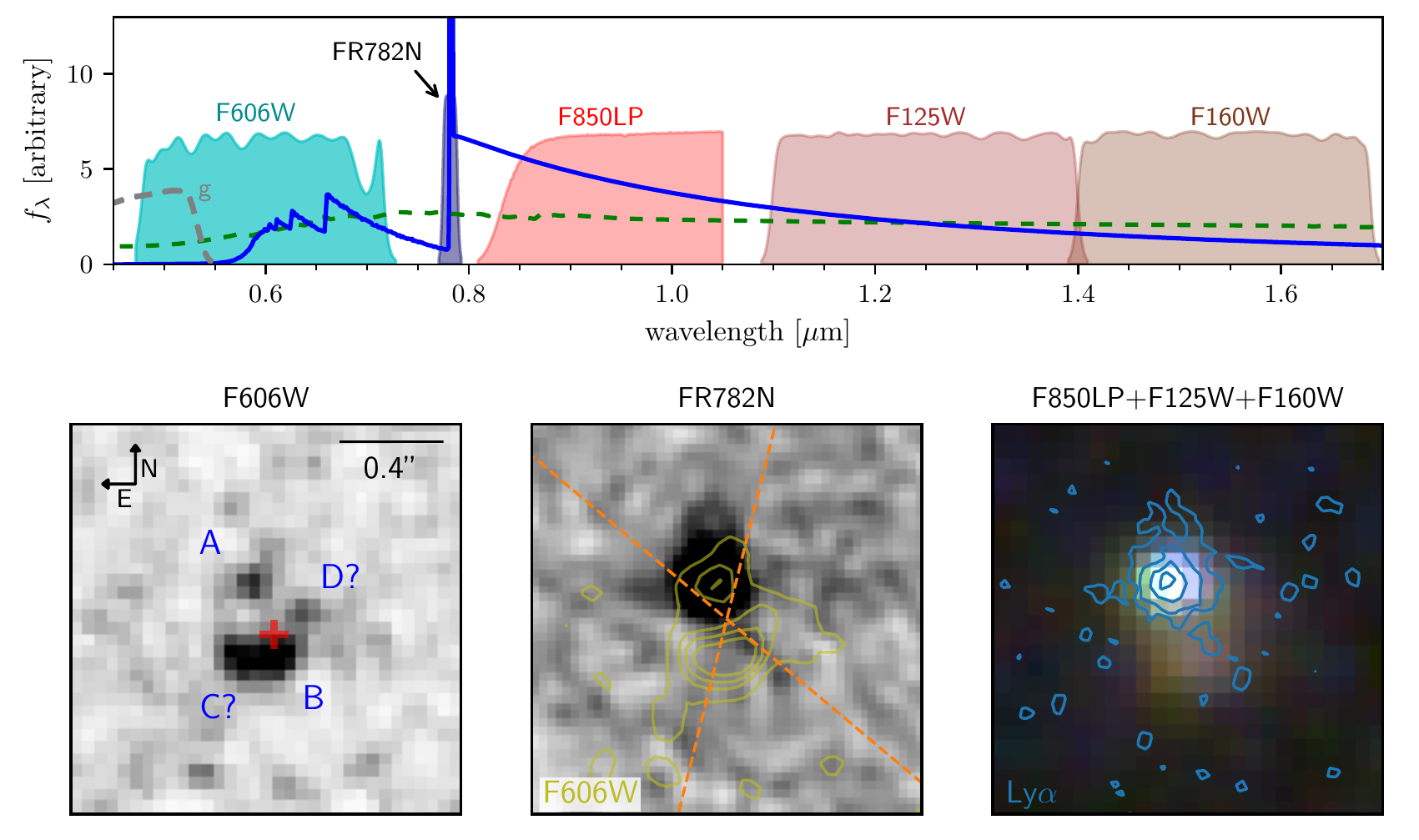}
\caption{HST imaging of \lae. The top panel displays the filter set, 
 including the ACS ramp filter used for narrow-band imaging of Ly$\alpha$
 (FR782N). A model for the LAE spectrum based on its observed properties
 is represented by a blue line, while an Sbc template for the foreground
 galaxy is represented by a green dashed line. The bottom three panels
 present the HST images, oriented as indicated in the lower left panel.
 The F606W image clearly has multiple components.
 The labels match those in Table~\ref{tab:hstfit}. Labels C and D
 mark low-significance peaks in the HST image that are putative detections;
 in the analysis C is modeled as both a separate component and as part of
 B, while D is ignored.
 The red plus sign marks the location of the $g$-band centroid obtained from
 the CFHT/LBT imaging. The middle panel presents the narrow-band image, 
 smoothed with a 1-pixel Gaussian and overlaid with linearly spaced contours
 from the F606W image. The approximate locations of the spectroscopic
 slits for the two MODS masks are indicated with dashed orange lines.
 The right panel presents a colour composite constructed from the ACS and
 WFC3 images, where the ACS image has been convolved with a Gaussian and
 resampled to match the WFC3 PSF and footprint. Logarithmically spaced contours 
 from the narrowband (Ly$\alpha$) image are overlaid in cyan lines.
 \label{fig:hstims}
 }
\end{figure*}

We generated empirical PSF models for the broadband images from stars within 
the field using IRAF DAOPHOT tasks. The narrowband image has only a single
stellar object with sufficient signal-to-noise to serve as a PSF reference.
\lae\ is well detected in all HST images, including the narrowband
(Fig.~\ref{fig:hstims}). We will discuss multi-component fits to the HST
images in \S~\ref{sec:morphology}.

\subsection{Spitzer Cycle 9 Observations}\label{sec:obs_spitzer}

Mid-IR observations have a key role in characterizing high redshift galaxies
by providing constraints on the SED redward of the Balmer break. Compared to
rest-UV wavelengths, rest-frame optical observations better probe the  
star formation history, stellar mass, and dust extinction, and thus are
essential to forming a more complete picture of the stellar population.

\lae\ was observed with the IRAC camera on Spitzer during the Warm Mission
(GO \#90195, PI: McGreer). The observations consisted of a single
AOR with integrations in the 3.6$\mu$m and 4.5$\mu$m channels totalling 1800s 
in each channel using a standard dither pattern.
We estimate a point source sensitivity of $\sim1.5~\mu$Jy (5 $\sigma$) from 
both the 3.6$\mu$m and 4.5$\mu$m images. \lae\ is clearly detected in both 
images, with total fluxes of 
$f_{3.6} = (4.08 \pm 0.33)~ \mu$Jy and $f_{4.5} = (5.58 \pm 0.30)~ \mu$Jy
measured through aperture photometry (\texttt{SExtractor} MAG\_AUTO).

\section{Image Modelling}\label{sec:imagemodel}

\subsection{Foreground galaxy}\label{sec:foreground}

The $g$-band detections, and marginal $u$-band detection, are somewhat 
puzzling. Both bands are completely blueward of the Lyman Limit at this 
redshift (see Fig.~\ref{fig:hstims}) and thus will be subject to
Lyman continuum absorption. The comoving mean free 
path of Lyman Limit photons at $z=5.4$ is $\sim60~h_{70}^{-1}$~Mpc 
\citep{Worseck+14}; hence the probability that significant flux would transmit
through the intergalactic medium (IGM) from a high-redshift galaxy into these bands is exceedingly low. We used Monte 
Carlo simulations\footnote{The simulations are described in greater detail in 
\citet{McGreer+13} and are based on the forest model of \citet{WP11}, extended
to include the incidence of high-redshift Lyman Limit Systems from \citet{SC10}.} 
of Ly$\alpha$ forest transmission spectra at $z=5.4$ to examine the possibility 
that an IGM sightline would be sufficiently transparent to allow detectable
$g$-band flux from \lae. After generating a model galaxy spectrum based on 
the photometry at longer wavelengths (conservatively assuming a blue UV slope
of $\beta_\lambda=-1.5$ and an escape fraction of unity), we find that 
none of the 2000 simulated sightlines had $g<28$, compared to the measured 
fluxes of $\sim26.5$. We conclude that the flux at blue wavelengths must be 
due to a foreground interloper.

An interesting question is whether the interloper would have impeded  
selection of the high redshift galaxy according to standard colour selection
methods. The criteria employed for our quasar selection are very similar to 
those used to select ``dropout'' galaxies. As noted previously, we did not 
reject this object as the $g$-band photometry in the catalogs we used for 
selection reported a $<2.5\sigma$ detection. However, a stricter cut on the 
bluer bands may have rejected this object. In addition, the faint fluxes in 
the bluer bands are sufficient to affect photometric redshift estimation. The 
CFHTLS photometric redshift catalogs of \citet{Ilbert+06} and 
\citet{Coupon+09} place \lae\ at $z=0.75$, with a 68\% confidence interval of 
$0.68 < z < 0.89$. The fluxes in the blue optical bands result in this 
high-redshift galaxy having roughly similar colours to the red galaxies at the 
cluster redshift, and thus it could be easily overlooked by broadband colour
selection.

\begin{table*}
 \begin{center}
 \caption{Photometry obtained from multiband fitting to HST images, 
 with the exception of FR782N which is obtained from aperture
 photometry using \texttt{SExtractor} MAG\_AUTO.}
 \label{tab:hstfit}
 \begin{tabular}{lcccccc}
  \hline
component & ($\Delta(\alpha),\Delta(\delta)$) & F606W & FR782N & F850LP & F125W & F160W \\
\hline
\multicolumn{6}{|c|}{Two-component model} \\
 PSF(A) &  (+0.000,+0.000) & $  27.10 \pm 0.18$  & $  20.79 \pm 0.04$ &$  23.64 \pm 0.02$  & $  23.84 \pm 0.01$  & $  23.95 \pm 0.01$  \\
 Sers(B+C) &  (-0.103,-0.260) & $  25.30 \pm 0.10$  & $>24.6$ &$  24.43 \pm 0.17$  & $  23.94 \pm 0.02$  & $  23.70 \pm 0.02$  \\
\multicolumn{6}{|c|}{Three-component model} \\
 Sers(A) &  (+0.000,+0.000) & $  27.27 \pm 0.89$  & $  20.79 \pm 0.04$ &$  23.54 \pm 0.02$  & $  23.71 \pm 0.01$  & $  23.80 \pm 0.01$  \\
 Sers(B) &  (-0.142,-0.307) & $  26.47 \pm 0.53$  & $>24.6$ &$  25.27 \pm 0.78$  & $  24.43 \pm 0.23$  & $  24.37 \pm 0.40$  \\
 Sers(C) &  (+0.017,-0.313) & $  25.87 \pm 0.42$  & $>24.6$ &$  27.27 \pm 5.55$  & $  25.92 \pm 0.27$  & $  25.14 \pm 0.21$  \\
\hline
 \end{tabular}
 \end{center}
\end{table*}

\subsection{HST image decomposition}\label{sec:morphology}

Interpretation of the high resolution images from HST is not straightforward. 
The WFPC2 F606W image (Figure~\ref{fig:hstims}) shows the largest degree of 
apparent structure, with up to four individual emission peaks. On the other 
hand, the narrow band image has only a single prominent source. The F850LP 
image is dominated by a bright source that coincides with the narrowband 
detection, with the addition of a faint source to the SW. The WFC3 images 
appear to have two components of roughly equal strength, roughly 
matching the morphology of the F850LP image.

It is not obvious how to associate individual components across all of the  
emission bands. The narrow band image pinpoints the location of the 
Ly$\alpha$ emission from the $z=5.4$ galaxy. Emission peaks at this position 
are present in all bands, including F606W (see the component labeled ``A'' in 
Figure~\ref{fig:hstims}). On the other hand, the foreground galaxy should 
dominate the $g$-band emission. The centroid obtained from the ground-based 
CFHT/LBT $g$-band imaging most closely aligns with the brightest component 
found in the F606W image (labeled B in Fig.~\ref{fig:hstims}). This position  
is also well matched to the extended emission to the SW of the Ly$\alpha$
peak in the HST images.

We implemented a multi-band fitting procedure in python that allows for an 
abritrary number of point source (hereafter PSF) and extended object 
components to be included in a given model. These components are then 
rendered into images for each HST band by convolving the model with 
empirical PSFs derived from field stars. The extended components are based on 
S\'ersic profiles. The rendered images are then compared to the data with a 
$\chi^2$ statistic, summing the contributions from all bands. A minimization 
routine (the Nelder-Mead gradient search implemented in Scipy) is then used to 
find the best-fit set of parameters. During this procedure we exclude the 
narrow band image as it probes only the Ly$\alpha$ emission from the high-$z$ 
galaxy, which may have a different morphology than the continuum emission.

We then experimented with a number of configurations to fit the individual 
emission components in the HST images. Unresolved components are 
modelled with three parameters ($x$, $y$, and flux), while S{\'e}rsic profiles
have a total of seven parameters ($x$, $y$, flux, effective radius, S{\'e}rsic
index, ellipticity, and position angle). Given the large number of parameters
and the substantial blending apparent in the images (leading to 
degeneracies in the fits), we reduce the number of parameters by making some 
simplifying assumptions. We fixed the position of the Ly$\alpha$-emitting 
component (A) to the position obtained from the F850LP image, where it is 
cleanly detected at high significance. We required the positions of any 
additional S{\'e}rsic components to be identical across all bands. Finally, we 
found that fitting the sky background was unneccessary and thus simply removed a 
median value from each band using nearby sky pixels.

\begin{figure}
\centering
\epsscale{1.0}
\plotone{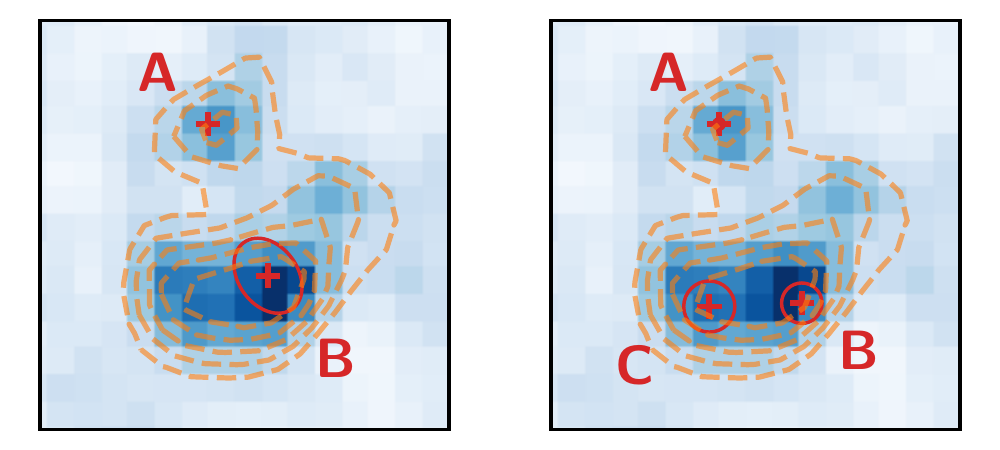}
\caption{Image models for the two (left panel) and three (right panel) 
 component fits to the HST images. The background image is from WFPC2/F606W,
 with a pixel scale of 0.05\arcsec. The dashed lines are linearly spaced 
 surface brightness contours from the F606W image. Positions of the
 individual components are labeled and marked with plus signs. Extended
 components (S{\'e}rsic profiles) are represented with an ellipse based
 on the fitted parameters (displayed without PSF convolution), with a radius 
 of $r_{\rm eff}/5$. The size of the image cutouts is 0.4\arcsec\ and the
 orientation is N through E, as in Figure~\ref{fig:hstims}.
 \label{fig:imagemodels}
 }
\end{figure}

Our fiducial configuration (left panel of Figure~\ref{fig:imagemodels})
consists of a single PSF component (A) aligned with the bright source in the 
F850LP image, and a S\'ersic component (B). We required the 
position angle of the S\'ersic component to be identical in all bands, and 
fixed the index to $n=1.1$ and the ellipticity to 0.28 in all bands. The 
latter values were found by allowing those two parameters to vary while 
holding other parameters fixed. The effective radius is allowed to vary 
between bands in order to account for morphological variations with wavelength.
This model provides a reasonably good fit to 
the data, with $\chi^2_\nu = 1.22~(16377/13426)$. The best-fit effective 
radius is $0.7$\arcsec\ $\pm 0.1$\arcsec\ in the F606W band and 
$0.9$\arcsec\ $\pm 0.03$\arcsec\ in 
the other bands. In this model we assume that the PSF component corresponds to 
all of the flux from the high-$z$ galaxy, and the S\'ersic component is 
contributed entirely by a low-$z$ interloper.

In the F606W image the SW component is highly asymmetric, with a strong
peak and extended emission to the east. We consider the possibility that
this emission represents a separate component by constructing a model
with three components, where all components are represented by S\'ersic 
profiles (right panel of Figure~\ref{fig:imagemodels}). 
The initial positions and fluxes were obtained from simple Gaussian fits to the 
individual peaks in the F606W image. We fixed the positions of all components 
to their initial positons, and 
also fixed the S\'ersic profiles to have zero ellipticity and required that 
the effective radii and S\'ersic indices were identical across all bands. 
Although the Ly$\alpha$-emitting component is modelled with a S\'ersic 
profile rather than a PSF as before, the best-fit effective radius is 
$\sim0.1$\arcsec, indicating that the source is at best marginally resolved in 
the broad-band images. Components B and C both have $r_e \approx 0.4$\arcsec\, 
and the best-fit S\'ersic indices are rather flat ($n\la$1). The fit is 
improved compared to the two-component model: 
$\chi^2_\nu = 1.20~(16060/13423)$.
The addition of three parameters is statistically significant according
to the Aikake Information Criteria ($\Delta{\rm AIC} \approx 300$).
However, splitting the fluxes between the B and C components results
in large photometric uncertainties, thus it is difficult to reliably
constrain the contribution from each component.

Furthermore, the three-component fit has more ambiguity in its interpretation. Associating 
component B with the foreground galaxy remains clear, but the second 
component (C) could either be in the foreground or at high-$z$; i.e., it could 
be continuum emission from the high-$z$ galaxy without associated Ly$\alpha$ 
emission, as is often observed \citep{Jiang+13,Pirzkal+07,Venemans+05}. We 
consider the latter interpretation to be less likely, as this would imply 
relatively bright emission in a very blue band ($V_{606} \approx 25.9$) along 
with relatively faint emission in the rest-UV continuum 
($J_{125} \approx 25.9$), compared to $V_{606} = 27.7$ and $Y_{125} = 23.7$ 
for the Ly$\alpha$-emitting component. In order to be more conservative we
adopt the two-component model to interpret the HST data.

\begin{figure}
\centering
\epsscale{1.0}
\plotone{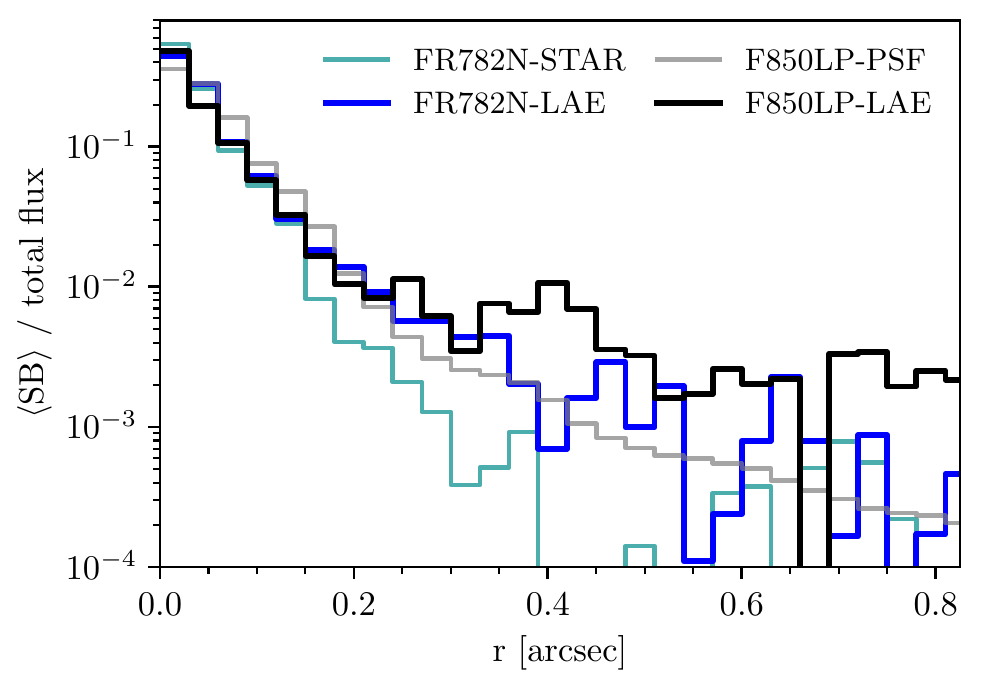}
\caption{Radial profiles of the Ly$\alpha$ (FR782N) and UV continuum
 (F850LP) emission, expressed as the average surface brightness of pixels
 within radial annuli of width 0{\farcs}03, normalized by the total flux
 in a 1.2\arcsec\ aperture. The profile from the PSF
 derived for the F850LP image is marked a light grey line, while the LAE
 profile is black. The profile from the single reference star in the
 narrowband image is shown in light blue, while the LAE profile is dark
 blue. Errors on the radial profiles are typically smaller than the
 line width and not displayed.
 \label{fig:radprofile}
 }
\end{figure}

\begin{figure*}
\epsscale{2.0}
\plotone{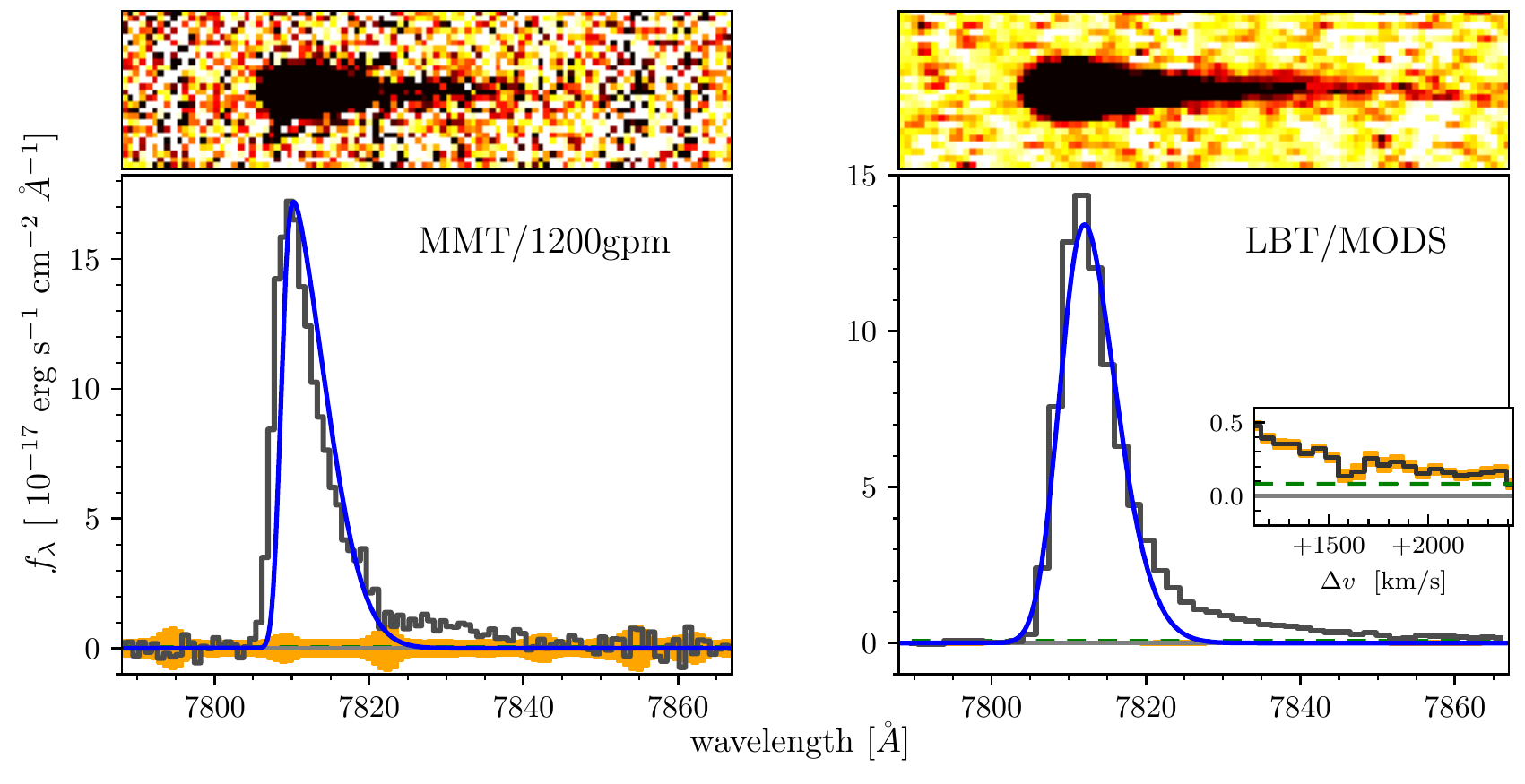}
\caption{Ly$\alpha$ emission profiles from the MMT/Red Channel 1200gpm 
 spectrum (left) and LBT/MODS (right). The upper panels display the 2D
 spectra over an extent of 8\arcsec\ for the MMT spectrum and 6\arcsec\ 
 for the LBT spectrum (the LBT spectrum has been binned by a factor of two
 along the spatial axis for display purposes).
 In the lower panels 
 the orange shaded regions span the $\pm1\sigma$ errors, and the green 
 dashed lines mark the continuum level obtained from fitting the spectrum
 at $\sim$8000\AA.
 The blue line shows the result of the truncated Gaussian profile fit, 
 convolved with the profile of each instrument. This profile fits the core
 of the line reasonably well, but fails to account for the extended red wing
 of the line.
 The inset panel displays a zoom on the MODS spectrum, showing that the 
 flux only drops to the
 continuum level at $\sim7870$\AA, corresponding to $+2300~\kms$ from the
 peak of the Ly$\alpha$ emission. 
 We compared the 2D profile of the high $S/N$ MODS spectrum with reference
 stars included in the same slit mask and see no evidence for spatially
 extended Ly$\alpha$ emission at $\sim1$\arcsec\ resolution.
 \label{fig:lyaprofile}
 }
\end{figure*}

\subsection{Narrowband Ly$\alpha$ Imaging}\label{sec:lyaim}

The ACS/FR782N band was selected to map the spatial extent of the Ly$\alpha$ 
emission from the $z=5.4$ galaxy with the resolution available from HST.
The bandwidth of the ramp filter is $\sim150$\AA, fully encompassing the
extent of the Ly$\alpha$ emission from spectroscopy ($\sim70$\AA, 
Fig.~\ref{fig:lyaprofile}). The narrowband image is presented in 
Figure~\ref{fig:hstims}. As discussed in the previous section, the single
narrowband detection is also well detected in the broad bands redward of
the Ly$\alpha$ line. Fitting a power law to the F850LP, F125W, and F160W
measurements (Table~\ref{tab:hstfit}) we obtain a slope of 
$\beta_\lambda = -2.58 \pm 0.03$.

The total flux in the narrow band is $m_{782} = 20.79 \pm 0.04$ as measured 
through a 0{\farcs}8 aperture. Extrapolating the continuum fit to this 
wavelength results in EW$_0$ = $260 \pm 12$\AA. One of the key aims of the 
HST program was to obtain resolved photometry of the rest-UV emission, 
removing the foreground contamination. As described in \S\ref{sec:uvspec}, 
the EW of the Ly$\alpha$ emission obtained from ground-based spectroscopy is
consistent with the value from the HST narrowband imaging once the foreground 
contamination is taken into account. Thus the large EW is robust, and is 
slightly greater than the limit generally assumed for emission from normal 
stellar populations \citep[240\AA, e.g.,][]{CF93,Schaerer02}, although
objects with similarly large Ly$\alpha$ EWs have been discovered in 
large narrowband surveys \citep[e.g.,][]{Hashimoto+17,Shibuya+17}.

The narrowband image contains a single stellar object with sufficient 
$S/N$ to characterize the PSF. Comparing the radial profile of this object
to the \lae\ detection, the latter is clearly more extended 
(Fig.~\ref{fig:radprofile}). For the star,
90\% of the total flux is contained within a 0{\farcs}15 radius, while for
the LAE this radius is 0{\farcs}45. The Ly$\alpha$ flux extends to 0{\farcs}6
in this relatively shallow image (the surface brightness limit is $\sim22.3$ 
mag~arcsec$^{-2}$). Note that component B is separated from the LAE
by $\sim$0.4\arcsec\ and contributes a small amount of flux to the F850LP
profile; however, it is undetected in the narrow-band image.
Several faint features appear to extend outwards from 
the central source after smoothing the narrowband image, although they are 
weak and do not permit a detailed morphological analysis.

In general, the lack of (or very weak) extended Ly$\alpha$ emission 
is in contrast to some recent work that finds extended Ly$\alpha$ halos
around high-redshift galaxies with strong Ly$\alpha$ emission, with
the line emission extending over a region 5--10$\times$\ greater than
the continuum emission \citep[e.g.,][]{Wisotzki+16,Smit+17}. \lae\ 
may lack the conditions required for such a halo to form. Alternatively,
the central region may be in a region of higher lensing magnification
compared to the more extended halo (see \S\ref{sec:lensing} for discussion
of the lensing properties).

\section{Physical Properties of the Galaxy}\label{sec:discuss}

\subsection{Spatial Extent}\label{sec:spatialextent}

From the analysis of the HST imaging presented in \S\ref{sec:imagemodel} 
we conclude that the both the UV continuum and Ly$\alpha$ emission from 
\lae\ are at best marginally resolved. Both show slight excess emission 
out to $\sim$0{\farcs}6 (Fig.~\ref{fig:radprofile}); however, it is difficult 
to draw robust conclusions on any differences between the UV continuum and 
Ly$\alpha$ morphology from the available images.

\subsection{Rest-frame UV spectra}\label{sec:uvspec}

\lae\ is exceptionally bright at optical wavelengths and provides a unique
opportunity to probe the physical conditions in a high-redshift galaxy using 
typical diagnostic tools applied to rest-frame UV spectra. The $\sim$9 hour
LBT spectrum achieves a $S/N$ of $\sim3$ in the rest-UV continuum ($\sim1250$\AA)
and can be used to explore key emission and absorption features at these 
wavelengths.

The foreground interloper complicates the interpretation of the
LBT spectrum. Fortunately, the HST imaging constrains the expected UV continuum 
from the LAE such that we can roughly correct for the foreground contamination.
This issue is evident at the stage of combining the spectra from the three
different nights: two of the masks of the MODS masks (\#505919 and \#510122) 
were aligned at a position angle (PA) of $-14^\circ$, nearly orthogonal to the
orientation between the foreground object and the LAE in the WFC3 images (see 
Fig.~\ref{fig:hstims}). The third mask (\#523405) was aligned at PA 
$= 50^\circ$, closer to parallel between the orientation of the two components.
A greater degree of foreground contamination would be expected in this mask,
and indeed, the EW of Ly$\alpha$ measured from this spectrum -- using a direct
fit to the observed continuum -- is nearly half that measured from the other two
spectra. We note there is no evidence for spatially extended emission in the 2D 
spectra at the resolution of the LBT seeing ($\sim$0.6-1.0\arcsec).

The full LBT/MODS1 spectrum is presented in Figure~\ref{fig:modsspec}, while
a zoom on the Ly$\alpha$ feature in both the LBT and MMT/1200gpm spectra can be 
seen in Figure~\ref{fig:lyaprofile}. Results obtained from analysis of the spectra
are given in Table~\ref{tab:specfit}.

\textit{Ly$\alpha$ emission:}
We first examine the strong Ly$\alpha$ emission feature. We extract a line
flux by integrating the spectrum between 7800\AA\ and 7850\AA\ 
after subtracting a fit to the continuum redward of the line at 8000\AA.
This observed flux -- $11.5 \times 10^{-16}~\fcgs$ -- is exceptionally large,
even when compared to other lensed galaxies at high redshift.
Ignoring gravitational lensing\footnote{The lensing correction is highly
uncertain and will be discussed in \S\ref{sec:lensing}.}, this corresponds 
to a line luminosity of $L_{{\rm Ly}\alpha} \sim 4 \times10^{44}~\ergss$.
We obtain a raw EW$_0$(Ly$\alpha$) $= 214 \pm 5$\AA\ using the continuum fit.
We estimate the contaminating flux of the foreground object to the continuum
using the HST photometry; after removing this flux we obtain a corrected value
of EW$_0$(Ly$\alpha$) $\sim 260$\AA, in excellent agreement with the results
from the HST narrowband image (\S\ref{sec:lyaim}). 
We also derive a  simple non-parametric estimate for the linewidth by measuring 
the red Half-Width at Half-Maximum (rHWHM; the width of the red side
of the line profile at half of the peak value). Using the higher resolution  
MMT/1200gpm spectrum and correcting for the instrumental profile 
(FWHM~$\sim100~\kms$) the measured rHWHM is $160~\kms$.

\begin{table}
 \begin{center}
 \caption{Summary of physical quantities obtained from analysis of the LBT
  spectrum and HST imaging.}
 \label{tab:specfit}
 \begin{tabular}{ll}
 \hline
 \hline
   Property  &  Value \\
 \hline
 R.A. (J2000) & 14:14:46.827 \\
 Decl. (J2000) & +54:46:31.94 \\
 $z$(Ly$\alpha$)       & 5.4253 (from peak wavelength)                 \\
 rHWHM(Ly$\alpha$)     & 160 $\kms$ (from MMT/1200gpm)                \\
 flux(Ly$\alpha$)      & ($11.5\pm0.3) \times 10^{-16}~\fcgs$       \\
 EW$_0$(Ly$\alpha$)    & $214 \pm 5$~\AA~~($\sim$260~\AA\ from phot.)                                          \\
 ${}^{(\star)}L_{{\rm Ly}\alpha}$  & $(3.8 \pm 0.1) \times 10^{44}~\ergss$     \\
 ${}^{(\star)}$SFR(Ly$\alpha)$       & $390 \pm 10~\Msunyr$                         \\
 \hline
 $z$(\ion{N}{IV}] 1486)    & 5.4237                                    \\
 FWHM(\ion{N}{IV}] 1486)    & $344 \pm 26 \ \kms$                     \\
 flux(\ion{N}{IV}] 1486)   & ($3.8\pm0.3) \times 10^{-17}~\fcgs$   \\
 EW$_0$(\ion{N}{IV}] 1486) & $7 \pm 3$~\AA~~(11~\AA\ from phot.)                                      \\
 $\Delta{v}$(Ly$\alpha$-\ion{N}{IV}])  & $+72~\pm~13~\kms$           \\
 \hline
 EW$_0$(\ion{N}{V})    & $<0.24$~\AA~~($<$0.5~\AA\ from phot.)                                          \\
 EW$_0$(\ion{C}{IV})   & $\la17$~\AA~~($\la$27~\AA\ from phot.)                                          \\
 ${}^{(\star)}M_{1350}$    & $-22.99$ (from phot.)                         \\
 \hline
 \end{tabular}
 \end{center}
 Notes: Rest-frame EWs are given in units of \AA, 
 line fluxes in $\fcgs$, continuum flux densities in $\flam$,
 luminosities in $\ergss$, and SFRs in $\Msunyr$.
 The reported position is from the HST F850LP image.
 The redshift estimate is obtained from the pixel wavelength corresponding
 to the peak flux density.
 Values indicated as being derived from photometry are obtained from the
 continuum fit to the resolved HST photometry. EW errors include a factor
 of 50\% uncertainty on the continuum level.\\
 $(\star)$ No lensing correction has been applied to the luminosities or quantities 
 derived from them.
\end{table}

It can be seen from Figure~\ref{fig:lyaprofile} that the Ly$\alpha$ emission 
extends to $>2000~\kms$ redward of the peak. This can be compared to a 
$z=5.7$ LAE reported by \citet{Yang+14} to have a red wing extending 
$>1000~\kms$ from the peak of the Ly$\alpha$ emission, and similar 
``shoulders'' observed in the red wings of bright, high-redshift LAEs 
\citep[e.g.,][]{Lidman+12,Smit+17}.
As with these examples, \lae\ is sufficiently bright --- both in the 
continuum and line emission --- to permit detailed studies of the line profile.

The observed features of the Ly$\alpha$ spectrum can be attributed to scattering 
through an outflow, which is usually represented as an expanding, dusty shell of 
neutral hydrogen \citep{Verhamme+08,GBD15}. We apply the automated fitting 
routine described in \citet{GBD15} to find the best-fit shell model for the
observed Ly$\alpha$ profile. Briefly, this method fits a model with six shell
model parameters to the observed spectrum: expansion velocity ($v_{\rm exp}$), 
hydrogen column density ($N_{\rm HI}$), effective temperature ($T$), intrinsic 
dispersion of the Ly$\alpha$ line ($\sigma_{\rm i}$), dust optical depth 
($\tau_{\rm d}$), and the intrinsic equivalent width of the line (EW$_{\rm i}$). 
In addition, the systemic redshift ($z$) is included in the fit. The best 
fit is determined through a $\chi^2$-minimization and the likelihood surfaces
are characterized with a Markov Chain Monte Carlo (MCMC) approach. Further 
details of the fitting techinique, including the parameter space explored by 
the simulations, may be found in \citet{GBD15}.

\begin{figure}
\centering
\epsscale{1.0}
\plotone{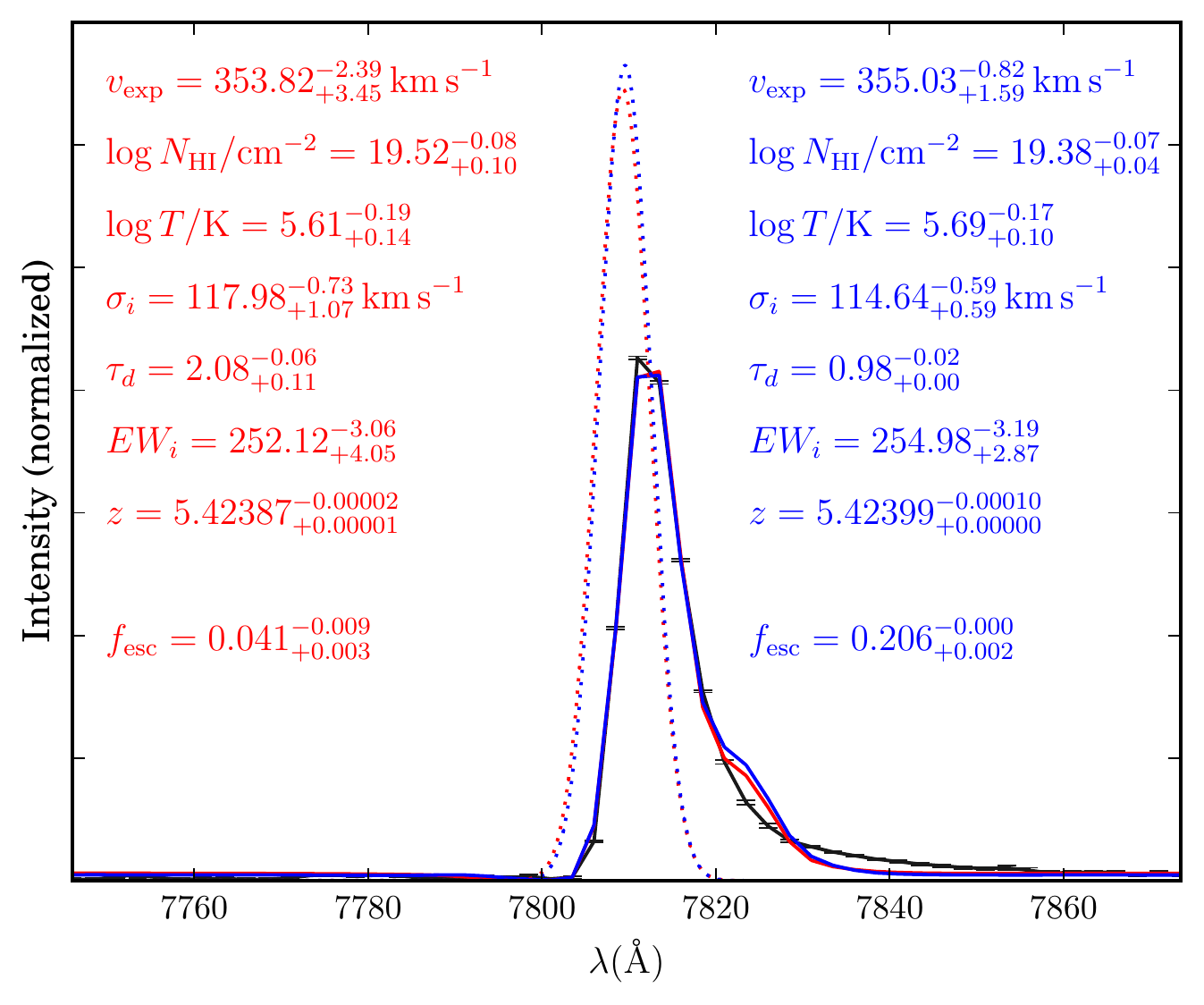}
\caption{Results of shell model fits to the Ly$\alpha$ line using the automated
 procedure from \citet{GBD15}. The black line shows the observed spectrum, 
 with horizontal error bars overlaid. The dashed lines show the intrinsic
 Ly$\alpha$ profiles from the model fits; the model with a large $\tau_{\rm d}$
 is shown in red and the model with a strong dust prior is shown in blue.
 The solid lines show the resulting profile after including radiative transfer
 through the expanding shell. Parameter values for the fit with large  
 $\tau_{\rm d}$ are given on the left and for the constrained $\tau_{\rm d}$
 on the right.
 \label{fig:lyafit}
 }
\end{figure}

Figure~\ref{fig:lyafit} presents the results of the shell model fits. The 
{\it solid red line} represents the best-fit model obtained after imposing a 
narrow Gaussian prior with $\sigma_v=15~\kms$ and centred at $z=5.4237$.
These values were selected based on the results of fitting the \ion{N}{IV}] 
line, which we assume to be at the systemic redshift. The data requires 
$(v_{\rm exp}, \log N_{\rm HI})\sim (350~\kms, 19.5)$. 
The observed Ly$\alpha$ line shift of $\Delta v_{{\rm Ly}\alpha}\sim 70~\kms$ 
primarily sets the constraint on $v_{\rm exp}$. The intrinsic width of the 
Ly$\alpha$ line is found to be FWHM$_{\rm int}\sim 250~\kms$. The latter is in remarkably good agreement with the width inferred from the \ion{N}{IV} line (see
below). What is surprising is that the best fit model favours the shell to be 
dusty, with $\tau_{\rm d} \sim 2$. This translates to Ly$\alpha$ escape fractions 
of $f_{\rm esc} \sim 10\%$. Given the large EW of the Ly$\alpha$ emission from
this galaxy, this low Ly$\alpha$ escape fraction is not likely 
to be physical. The {\it solid blue line} shows the best-fit model, if we force 
a lower $\tau_{\rm d}$ through a strong prior and also fix the redshift at
$z=5.24$. The data 
still favours dusty solutions with $\tau_{\rm d} \sim 1.0$ (the other model 
parameters are barely affected), leading to a Ly$\alpha$ escape fraction of 
$f_{\rm esc} \sim 30\%$. 
This is still low, but at least consistent with inferred Ly$\alpha$ escape 
fractions for galaxies that have similar $\Delta v_{{\rm Ly}\alpha}$ 
\citep[e.g.,][]{Erb+14,Yang+17}.
It is not possible to find shell model 
solutions with lower $\tau_{\rm d}$: the observed velocity shift of the 
Ly$\alpha$ line constrains the parameters $(v_{\rm exp}, \log N_{\rm HI})$. 
Generally, Ly$\alpha$ scattering broadens spectral lines in the absence of dust. 
In the presence of dust, scattering and spectral broadening are limited. The 
observed narrowness of the Ly$\alpha$ spectral line (and the absence of 
a blue peak) then requires at least some dust. Another possibility to 
be considered at this high redshift is the impact of the IGM which also narrows 
the observed Ly$\alpha$ line and extinguishes the blue peak 
\citep{Dijkstra2007_IGM}, thus, mimicking the effect of dust 
\citep[as discussed in][]{Gronke2017}.

The simple shell model has difficulty reproducing both the observed shift 
of the line ($\Delta v \sim 70$ km s$^{-1}$) and the very extended red wing. 
With a narrow prior on redshift, the model fails to reproduce the observations 
at $\lambda > 7825$\AA\ (corresponding to $\Delta v > 500~\kms$). If a wide 
redshift prior is employed, the model is able to reproduce the extended red 
wing, but underpredicts the overall shift of the line. This likely represents 
a shortcoming of the shell model. This tension could be alleviated by 
introducing a velocity gradient in the shell that would tend to ``smear out''
the observed spectrum \citep[e.g.,][]{LR99}.

\textbf{\textit{\ion{N}{V} emission:}}
There is no \ion{N}{V} $\lambda1240$ emission as would be expected from an 
AGN. The continuum is detected at $\sim4\sigma~{\rm pix}^{-1}$ in the 
region where \ion{N}{V} would be located (and has relatively little foreground
contamination); combined with the extremely large flux of the Ly$\alpha$ line
 we obtain a stringent constraint of 
$f$(\ion{N}{V})/$f$(Ly$\alpha$) $< 10^{-3}~(1\sigma)$, compared 
to $\sim$10\% for narrow-line AGN \citep{Alexandroff+13}.

\begin{figure}
\centering
\epsscale{1.0}
\plotone{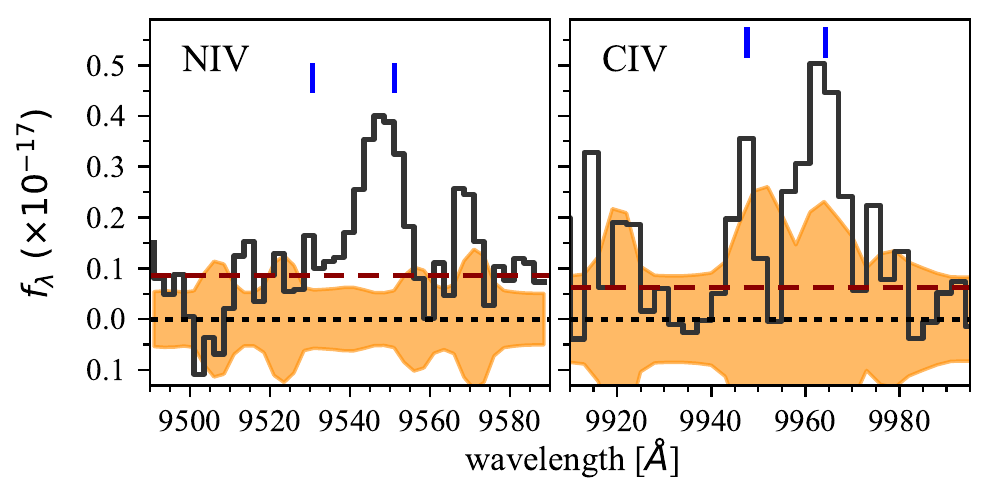}
\caption{Zoom on the LBT spectrum at the wavelengths of the \ion{N}{IV}] and 
\ion{C}{IV} emission lines. The spectral flux density is represented by the
 solid lines, and the $\pm1\sigma$ noise level by the orange shaded region.
 The continuum level is denoted with a red dashed line.
 The blue vertical lines mark the expected locations of the doublet features for
 both lines using the redshift obtained from the peak of the Ly$\alpha$ line.
 \label{fig:metallines}
 }
\end{figure}

\textbf{\textit{\ion{N}{IV}] emission:}}
The only other clear detection of an emission line in the LBT spectrum is
at 9549\AA. We identify this line as \ion{N}{IV}] 1486.5. While this emission
line is rarely observed, it has been found in a galaxy with very similar
properties to the one presented here. \gdssrc\ at $z=5.56$ has 
EW(\ion{N}{IV}] 1486.5) $\approx 30$\AA\ \citep{Vanzella+10}. This emission 
feature is a doublet, but in both cases no emission corresponding to 
\ion{N}{IV}] 1483.3 is detected (we obtain a limit of 
$f$(\ion{N}{IV}] 1483) $< 1 \times 10^{-17}~\fcgs$). 
The width of the line is ${\rm FWHM}~= 260 \pm 40~\kms$ after correcting for
instrumental resolution. The \ion{N}{IV}] emission
feature is displayed in Figure~\ref{fig:metallines}.

\textbf{\textit{\ion{C}{IV} emission:}}
The expected wavelength of 
\ion{C}{IV} $\lambda1550$ is at 9950\AA. Unfortunately, this is in a spectral
region with strong night sky emission lines.  The spectrum obtained from 
\texttt{modsIDL} processing appears to have significant positive flux in this 
region (see Fig.~\ref{fig:modsspec}), but the sky subtraction in this region 
did not seem to be reliable. We reprocessed the 2D spectra in this region 
using a custom 2D spline fit to the background sky emission. The sky subtraction 
residuals improved considerably after this reprocessing and the residual flux 
decreased; however, a small positive flux remains. Figure~\ref{fig:metallines} 
displays the reprocessed spectrum. Although the postive flux residual lies almost
exactly at the wavelengths expected for the \ion{C}{IV} doublet, these wavelengths
also align closely with two particularly strong night sky lines and it is difficult
to assign a significance to the flux. Taking the noise model for the sky background
at face value, we obtain a flux of $f$(\ion{C}{IV}) $\la 6 \times 10^{-17}~\fcgs$.
This flux is well below expectation for an AGN, with a flux ratio 
$f$(\ion{C}{IV})/$f$(Ly$\alpha$) $\la 0.06$, compared to
$\sim 0.2\mbox{--}0.5$ for narrow-line AGN \citep{Alexandroff+13,FO86}. 

\textbf{\textit{ISM absorption features:}}
Given the bright continuum flux of \lae\ we are able to obtain constraints
of faint interstellar medium (ISM) absorption and emission features. These features
provide important diagnostics of the kinematics and covering fraction of neutral
gas in the ISM. Stacking analyses provide the best constraints on the
strengths of these features in typical $z>3$ LBGs \citep[e.g.,][]{JSE12}; however,
a few such galaxies are sufficiently bright for individual study
\citep[e.g.,][]{Christensen+12,Bayliss+14,Patricio+16}.
In our case, the deep LBT spectrum of a $z=5.4$ galaxy only achieves 
$S/N \sim 3~{\rm pix}^{-1}$ at $\ga1200$\AA, and the foreground contamination
further reduces the sensitivity to faint spectral features. Nonetheless, we 
examine several of the ISM features previously studied by \citet{JSE12} in stacked
$3<z<7$ LBGs, using the wavelengths and velocity offsets listed in their Table~1,
and adopting Gaussian profiles with widths tuned to match their composite spectrum.

No ISM features are clearly detected in the LBT spectrum. 
\ion{Si}{II}~$\lambda$1260 has a marginal detection, with 
$W = -0.5 \pm 0.4~(1\sigma)$.
The \ion{Si}{II}$^*$~$\lambda$1264 fine structure line is also marginally detected,
with $W = 0.3 \pm 0.2$, as well as \ion{Si}{II}$^*$~$\lambda$1309 with
$W = 0.5 \pm 0.3$.
We obtain limits of $W<0.4~$\AA\ on the \ion{O}{I}+\ion{Si}{II}~$\lambda$1303
feature, and $W<0.2~$\AA\ on \ion{C}{II}~$\lambda$1334.
All other features are too strongly affected by night sky lines to produce
meaningful constraints. We have not attempted to correct for the foreground
contamination in these measurements; they mainly serve as a guide as to what
level of constraint we can obtain given the quality of the available spectrum.

In general the lack of strong ISM absorption features is consistent with the
trend that the equivalent widths of such features anticorrelate with the strength
of the Ly$\alpha$ emission, as detailed in \citet{JSE12}. Thus we would not
have expected to see ISM features in the spectrum of \lae. However, given the 
strength of the observed continuum flux an even deeper spectrum obtained with a 
30m-class telescope would place stronger constraints on the physical conditions
of the ISM in this galaxy.

\subsection{Ionizing spectrum}

We argued in the previous subsection that \lae\ lacks the spectroscopic
signatures of AGN. However, the \ion{N}{IV}] detection and questionable
\ion{C}{IV} detection indicate the nebular gas is subjected to a hard
radiation field that is able to ionize these higher order species, with
ionization potentials of 47.4~eV and 47.9~eV, respectively. Nebular
\ion{C}{IV} detections have been reported in lensed galaxies at $z=7.045$ 
\citep{Stark+15civ}, $z=6.11$ \citep{Mainali+17}, and $z=4.88$ \citep{Smit+17}.
\ion{N}{IV}] has been found in a massive galaxy at $z=5.56$ \citep{Vanzella+10},
and in lensed galaxies at $z=3.4$--$3.5$ \citep{Fosbury+03,Patricio+16}. 
While in none of these galaxies --- including the one
reported here --- can an AGN contribution be definitively ruled out, 
photoionization modeling shows that a population of very hot, massive stars
could account for the observed nebular emission, in particular the detections
of of high-ionization species such as \ion{N}{IV}]/\ion{C}{IV} without 
\ion{N}{V} emission as expected from an AGN \citep[e.g.,][]{Fosbury+03}. 
These conditions may be more prevalent during the early stages of galaxy 
formation and bright sources like \lae\ demonstrate the promise of rest-UV 
spectroscopy for probing the physical conditions of distant galaxies.
It is also noteworthy that \ion{N}{IV}] is not detected in many galaxies 
with \ion{C}{III} and \ion{C}{IV} emission \citep[e.g.,][]{Stark+14},
indicating it may only arise in particular conditions requiring more than
just a hard radiation field.

\subsection{Ly$\alpha$ velocity offset}

The detection of UV metal lines further provides a measure of the systemic 
redshift of the galaxy \citep[e.g.,][]{Stark+15ciii}. The velocity offset of 
the Ly$\alpha$ line ($\Delta{v}_{{\rm Ly}\alpha}$) is sensitive to the covering 
fraction and kinematics of neutral hydrogen;
as discussed in section \ref{sec:uvspec} the Ly$\alpha$ profile of \lae\ 
is broadly consistent with an expanding, dusty shell of neutral gas.
Here we examine the velocity offset of Ly$\alpha$ relative to the \ion{N}{IV}]
emission line. The offset between the centroid of the \ion{N}{IV}] line and
the peak of the Ly$\alpha$ is 70~$\kms$. \citet{Stark+15ciii} measured 
$\Delta{v}_{{\rm Ly}\alpha}$ for two lensed $z\sim6$ galaxies with systemic
redshifts from \ion{C}{III}] nebular emission and found smaller offsets than 
galaxies at $2<z<3$; however, \citet{Willott+15} detected relatively large 
($>400~\kms$) offsets in two luminous $z\sim6$ galaxies with systemic redshifts 
from [\ion{C}{II}]$\lambda$158~$\mu$m and argued the observed offsets are 
consistent with the overall trend between $\Delta{v}_{{\rm Ly}\alpha}$ and 
EW$_{{\rm Ly}\alpha}$ seen at lower redshift.
The relatively small velocity offset found for \lae\ by adopting the 
\ion{N}{IV}] redshift is similar to the results of \citet{Smit+17}, who found
an offset $<100~\kms$ for a lensed $z=4.88$ galaxy with an extended Ly$\alpha$
halo (and similarly large Ly$\alpha$ EW). Larger velocity offsets aid the escape 
of Ly$\alpha$ photons through the high-redshift IGM \citep[e.g.,][]{Stark+16}, and 
thus galaxies with both small velocity offsets and large Ly$\alpha$ equivalent 
widths are highly intriguing.

\subsection{Lensing Model}\label{sec:lensing}

The lensing group \slsid\ was identified by the SL2S \citep{Cabanac+07} 
based on the detection of a tangential arc (T1) and a candidate bright 
radial arc (R1). \citet{More+12} subsequently identified an additional 
radial arc candidate (R2), bringing the total number of lensed galaxy 
candidates associated with \slsid\ to three. Images of the three 
candidate arcs can be viewed in Figure~\ref{fig:clusterim} in 
Appendix~\ref{sec:appendix}.

\citet{Cabanac+07} estimated a photometric redshift of $z=0.75$ for the 
galaxy group, while \citet{More+12} estimated $z_{\rm lens} = 0.63 \pm 0.02$. 
We included several candidate group member galaxies in our 
MODS mask; more complete details are provided in Appendix~\ref{sec:appendix}. 
We obtain a mean redshift of 
$\langle{z_{\rm lens}}\rangle=0.613$ from spectra of 43 galaxies in the 
group, in excellent agreement with the \citet{More+12} photometric 
redshift.

The lensing properties of \slsid\ have been discussed in a number of
publications. \citet{More+12} quote an arc radius of $14.7$\arcsec\ 
for the tangential arc T1. 
\citet{Foex+13} report a photometric redshift of 
$z_s=1.47^{+0.75}_{-0.53}$ for the arc, and used a shear profile
analysis to obtain a velocity dispersion of 
$\sigma_{\rm SIS} = 969^{+100}_{-130}~\kms$ for the lens, 
roughly at the boundary
between a group- and cluster-scale lensing mass. Recently, \citet{Gruen+14}
identified \slsid\ with a galaxy cluster detected by the Planck satellite
through the Sunyaev-Zel'dovich (SZ) effect \citep{Planck+14}. 
\citet{Gruen+14} update the results of \citet{Foex+13} and obtain a 
significantly larger velocity dispersion through weak lensing analysis,
$\sigma_{\rm SIS} = 1540^{+162}_{-190}~\kms$. Their measurement of the
halo mass is $M_{500c} \approx 20\times10^{14}h_{70}^{-1}~\Msun$
from weak lensing and $M_{500c} \approx 8\times10^{14}h_{70}^{-1}~\Msun$
from the SZ detection. These larger estimates favour a cluster scale for the 
foreground mass.

\citet{Gruen+14} note that a redshift of $z_s=1.49$ for the source galaxy
of the tangential arc is favoured by their mass model in order to match 
the observed arc radius, in agreement with the photometric redshift 
estimate from \citet{Foex+13}. Spectroscopic redshifts for the candidate
arcs are strongly constraining on the lens mass models and hence we 
observed all three arcs with MODS. For R1 and R2 we obtain redshifts that
place them in the foreground of the cluster. For T1 we do not detect any 
emission lines in the MODS spectra; however, the strongest available line 
would be [\ion{O}{II}] $\lambda$3727 which at $z\ga1.5$ would be at 
wavelengths dominated by OH sky lines and thus difficult to detect.
Further details of the MODS spectroscopy of targets in the field of \lae\ 
are given in Appendix~\ref{sec:appendix}.

\begin{table}
 \begin{center}
  \caption{Photometry for the foreground galaxy from CFHT and HST.}\label{fgphot}
  \begin{tabular}{lr}
  \hline
   Band & AB mag \\
  \hline
  $u$       & $26.45 \pm 0.28$ \\
  $g$       & $26.58 \pm 0.23$ \\
  $V_{606}$ & $25.30 \pm 0.10$ \\
  $z_{850}$ & $24.43 \pm 0.17$ \\
  $H_{160}$ & $23.70 \pm 0.02$
  \end{tabular}
 \end{center}
\end{table}

We first consider a rough estimate for the lensing magnification
of the LAE based on a few simplifying assumptions.
\lae\ is $\sim$30\arcsec\ from the apparent cluster centre based on 
the BCG position, and is thus at a radius of $\approx2\theta_{\rm E}$, where
$\theta_{\rm E}$ is the Einstein radius. As an example, if we assume
an arc redshift of $z\approx2$, the lensing magnification from
the cluster alone would be $\mu \approx 3$. However, additional 
magnification from the foreground galaxy must also be taken into account. 
Adopting the Faber-Jackson relation given in \citet{Bernardi+03} and assuming 
the foreground galaxy is at the cluster redshift ($z=0.6$) yields a velocity
dispersion $\sigma_v \approx 85~\kms$ and $\theta_{\rm E} = 0.2$\arcsec.
Although the Faber-Jackson relation may not be appropriate for this
galaxy, this rough estimate demonstrates that the LAE may lie near the
Einstein ring radius. Thus the magnification is difficult to constrain and
may be rather large.
The HST narrowband imaging provides strong evidence against 
multiple-image strong lensing of the Ly$\alpha$-emitting component.
The WFPC2 imaging of the far-UV continuum is more ambiguous in that the
multiple components are suggestive of a strong lensing configuration.
However, we argued in \S\ref{sec:morphology} that only a single component 
in the WFPC2 image is likely associated with the LAE based on the observed 
colours. Thus while some ambiguity remains (which could be resolved with
further HST or JWST observations), the available HST imaging disfavours 
multiple image strong lensing.

\begin{figure}
\centering
\epsscale{1.0}
\plotone{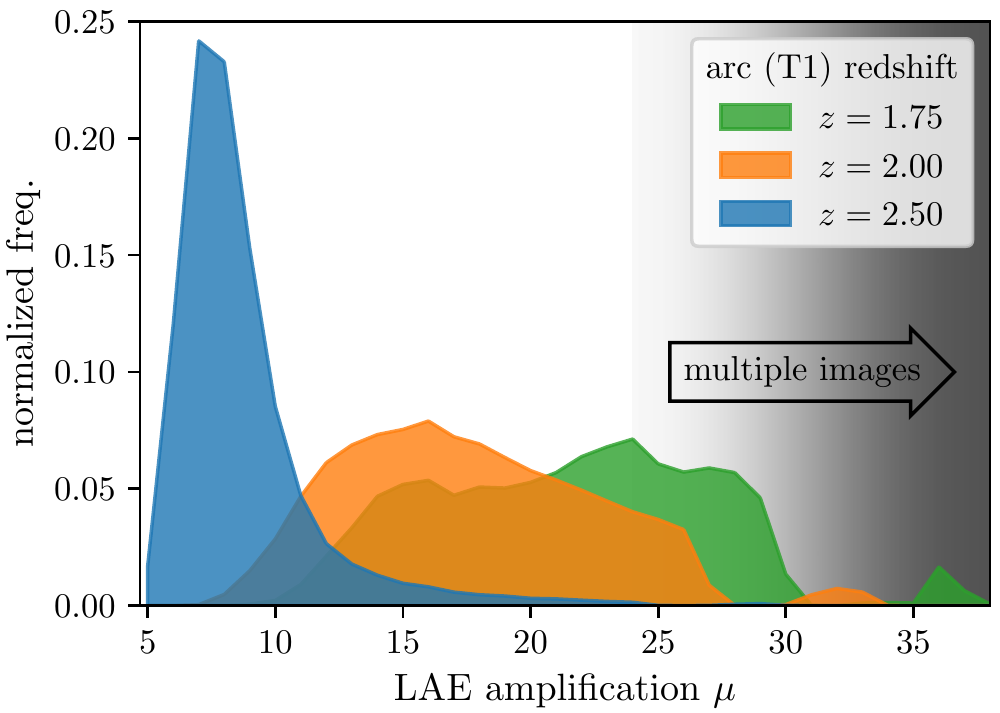}
\caption{Lensing amplification factors for \lae\ under three assumptions for
 the redshift of the lensed arc T1. The histograms are the result of Monte
 Carlo simulations accounting for uncertainties in the foreground interloper
 photometry and hence its estimated mass, assuming it lies at the redshift
 of the cluster. Simulations with $\mu \ga 25$ tend to result in multiple
 images and are unlikely given the observational constraints.
 \label{fig:lensingamp}
 }
\end{figure}

In order to better constrain the lensing magnification we model the
cluster mass through an MCMC approach using the Lenstool software 
\citep{Jullo+07}, which fits the normalization of the Faber-Jackson 
scaling relation. The input data are the cluster member galaxies with
redshifts from MODS spectroscopy and photometry from HST. The model 
includes the contribution from the foreground galaxy, which
can produce a wide range of amplifications depending on the assumed
mass. We thus fold in the photometric uncertainties on the foreground
galaxy, and make the simplifying assumption that it lies at the cluster
redshift.

We execute a series of Monte Carlo simulations that convolve the
uncertainties on both the normalization of the scaling relation 
and the foreground galaxy mass, excluding models that result in multiple
image lensing of the LAE (treated as a geometric point source) as the HST 
imaging suggests this is unlikely.
Finally, we assume a redshift for the lensed arc. The total amplification
increases with decreasing arc redshift; in fact, $z=1.5$ is essentially
ruled out as this invariably results in multiple images. In the HST imaging 
it is apparent that the lower part of the arc T1 is coincident with an 
unassociated source of roughly equal brightness. The two objects are blended 
in the CFHT imaging and this may have affected the previously reported 
photometric redshifts for the arc (i.e., biasing them to lower redshift).
We also note that the large mass estimates from \citet{Gruen+14} would 
produce multiple images of the LAE  at other locations in the cluster 
field, which we do not identify in the HST or CFHT images.

Compared to previous mass modeling of this system,
our approach has the advantage of self-consistently including the
new member galaxy redshifts we have obtained, the new constraints on 
the lensed arc candidate redshifts, and most crucially, the presence 
of the foreground galaxy near the LAE position which strongly perturbs
the local magnification map. This results in a highly non-linear 
estimate for the total lensing magnification of the LAE. Furthermore,
the unknown redshift for the lensed arc T1 adds considerable uncertainty 
to our magnification estimate. 
We thus investigate the impact of the arc T1 redshift on the LAE 
magnification by considering three discrete values of redshift in the 
range $1.5 < z \le 2.5$\footnote{The arc is unlikely to lie at $z>2.5$ 
given that it is detected in the $u$-band.} in order to explore the 
uncertainty of the magnification factor.

Figure~\ref{fig:lensingamp} presents the distribution of lensing
amplifications obtained from these simulations. For $z=2$ the 
median and interquartile ranges (IQR) are $\mu = 17^{+4}_{-3}$.
At $z=2.5$ this drops to $\mu = 8^{+1.4}_{-1.0}$. Given all the uncertainties 
involved in this analysis, we adopt $5 \la \mu \la 25$ as a conservative estimate 
for the range of allowed magnifications, thus the intrinsic luminosity of 
\lae\ is poorly constrained. It is likely
that the source is intrinsically round and experiencing a relatively
low magnification, although scenarios in which the intrinsic source
shape conspires with the lensing geometry to produce a high magnification
event with little apparent stretching cannot be ruled out. A spectroscopic
redshift for the lensed arc T1 would reduce the uncertainty on the
magnification but still allow a wide range of values. JWST will have
the capability to obtain resolved near-IR spectroscopy of both the 
foreground interloper and the LAE and thus provide robust constraints
on the lensing amplification of this high-redshift galaxy.

\subsection{Broadband SED}\label{sec:sedfit}

We further characterize the properties of \lae\ through a detailed analysis
of its SED. Before attempting to fit the SED, we must first deblend the 
foreground contribution from the observed fluxes, particularly in the long 
wavelength bands where the emission is completely unresolved. We approach this 
problem by constructing SED templates for the foreground galaxy that reproduce 
the (semi-)resolved photometry in the bluer bands, and use these templates to
account for the foreground contribution to the redder bands. This
compares to \S\ref{sec:morphology}, where we took advantage of the higher
resolution provided by HST to obtain deblended photometry of the foreground
and LAE  at $\lambda < 2\mu{m}$ (we consider this photometry to be 
"semi"-resolved in that the peaks of the two primary objects are well separated, 
but the light profiles are significantly blended). Here we attempt to use 
simple assumptions about the SED of the foreground galaxy to obtain constraints 
on the long-wavelength fluxes of the LAE.

\begin{figure}
\centering
\epsscale{1.0}
\plotone{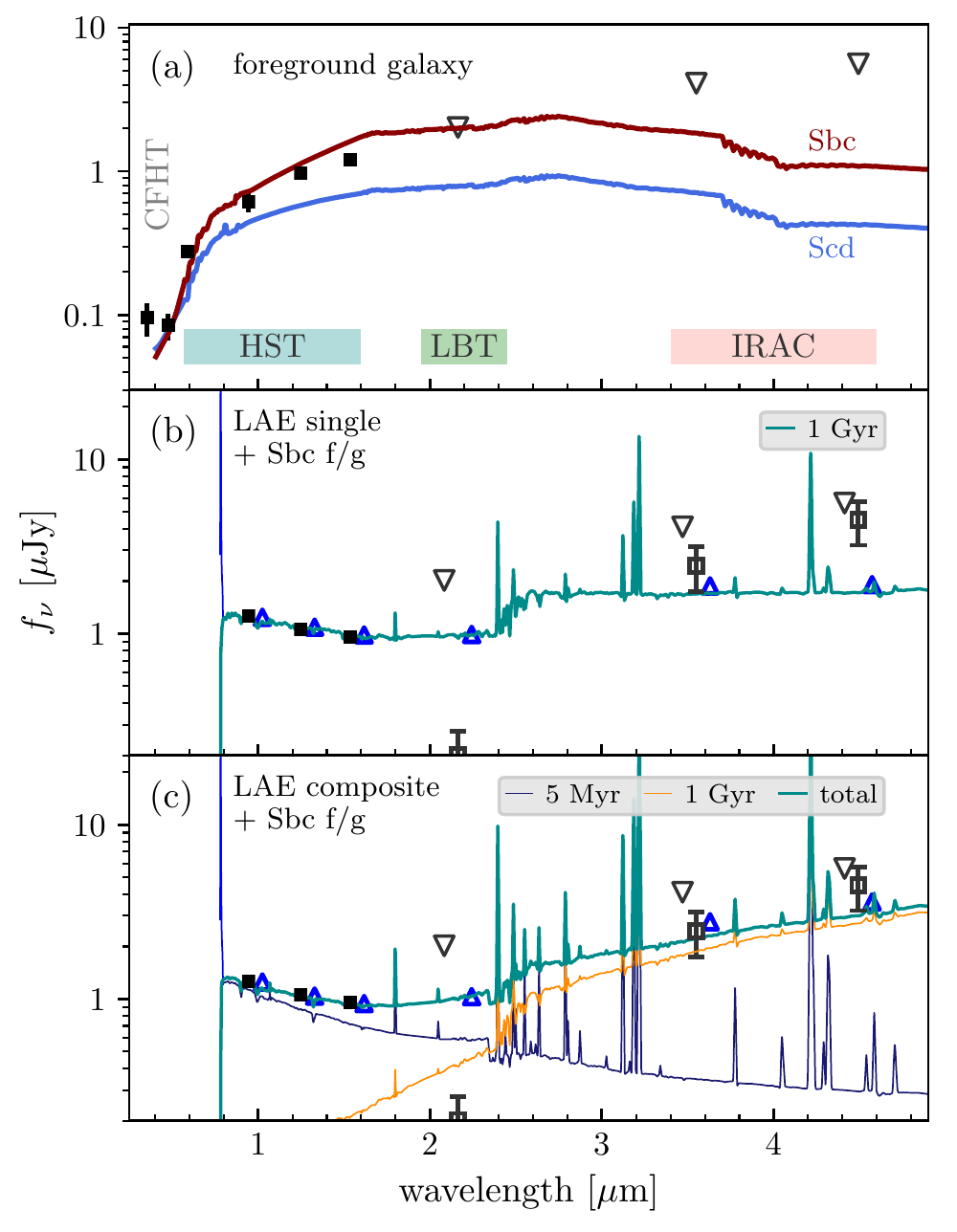}
\caption{Photometry and results from SED fitting.
 The black points with error bars represent the resolved CFHT and HST photometry.
 The inverted triangles mark the blended photometry from the LBT and Spitzer.
 Panel (a) shows SED models assumed for the foreground galaxy. 
 Based on the optical photometry the foreground SED is likely to lie between 
 the Sbc and Scd templates; this provides a range of values used for 
 subtracting the foreground contamination to the infrared data.
 Panels (b) and (c) present fits to the SED of the $z=5.4$ LAE. 
 The grey points with error bars are obtained subtracting the 
 foreground contamination assuming an Sbc SED. The fits are performed to these 
 ``decontaminated'' data. Panel (b) displays the fit for a single stellar 
 population model with a continuous star formation history, while panel (c)
 presents a composite model consisting of a young (5~Myr) starbursting
 component and an evolved (1~Gyr) population. The blue triangles mark the
 photometry recovered from the SED fits, offset slightly in wavelength.
 \label{fig:sed}
 }
\end{figure}

The $u$- and $g$-band detections are absent any contribution from the LAE and
those fluxes can be assigned to the foreground. The HST data provide photometry 
from 0.6$\mu$m--1.6$\mu$m, where we adopt the two-component model from
Table~\ref{tab:hstfit} and attribute the fluxes from B+C to the foreground.
Panel (a) in Figure~\ref{fig:sed} displays the resulting SED for the foreground
galaxy, where the points at $\lambda < 2\mu$m represent the (semi-)resolved
photometry.

In order to infer the contribution of the foreground to the total fluxes at 
$\lambda > 2\mu$m, we employ the template SEDs from \citet{CWW80}. 
These empirical templates are adopted as they provide a small
number of simple galaxy archetypes that we utilize to roughly span the
range of possible SEDs for the foreground galaxy. We
make the simplifying assumption that the foreground galaxy lies at the 
redshift of the \slsid\ group\footnote{The foreground galaxy is likely to be at 
low redshift given the marginal $u$-band detection.}. After fitting to the 
$\lambda < 2\mu$m data we find that an Sbc template represents the ``maximal'' 
contribution from the foreground galaxy that provides a reasonable fit to the 
resolved HST photometry. A bluer Scd template is a better match to the $u-g$ 
colour from the CFHT photometry but a poorer match to the HST data. Panel (a) 
of Figure~\ref{fig:sed} compares these templates to the observed SED of 
the foreground galaxy. The contributions to the total fluxes 
(Table~\ref{tab:phot}) in the longer 
wavelength bands for the Sbc (Scd) template are $\sim$90\% (40\%), 40\% (20\%), 
and 20\% (10\%) for the $Ks$, 3.6$\mu$m, and 4.5$\mu$m bands, 
respectively\footnote{We also examined photometry of field galaxies covered 
by the same imaging bands and found that those with SED shapes most similar 
to the foreground galaxy fell between the Sbc and Scd templates, bolstering 
the case for using these to bracket the possible foreground contamination.}.

Having characterized the SED of the foreground galaxy, we construct the SED for 
the $z=5.4$ LAE from the HST photometry, the LUCI $Ks$-band image, and the 
Spitzer/IRAC data. In total we have six data points spanning 1400\AA\ - 7000\AA\ 
in the rest frame of \lae. We construct two versions of the SED, alternately
using the Sbc or Scd template fits to subtract the foreground contamination 
to the $Ks$/3.6/4.5 bands. We further inflate the photometric errors in these 
bands by 0.3~mag in order to account for the uncertainty associated with the 
foreground removal. The Sbc foreground template is more conservative in the
sense that it results in smaller fluxes for the LAE in the IRAC bands; the
SED obtained after subtracting the Sbc foreground template is displayed in 
panels (b) and (c) of Figure~\ref{fig:sed}.

The deblended SED for the LAE is then fit with the method described in 
\citet{Stark+13}, which includes models for nebular emission presented 
in \citet{Robertson+10}. Briefly, the stellar population is
represented with templates from the \citep{BC03} models, and the nebular
emission is self-consistently included by using the ionizing photon output
from the stellar population.
Accounting for the nebular emission in this galaxy is
important as H$\alpha$ lies in the 4.6$\mu$m band and [\ion{O}{III}] lies in 
the 3.6$\mu$m band (H$\beta$ is just outside of this band). In all fits
we exclude the $Ks$ band, as it has the largest degree of uncertainty in
terms of the foreground contribution. The best-fit stellar template 
is found using a grid search of the stellar population parameters and 
identifying the template with the minimum $\chi^2$ calculated from the 
observed photometry. Dust extinction is included using the \citet{Calzetti}
relation.

The result of the single population fit is shown in Figure~\ref{fig:sed}(a).
We find that a single stellar population model provides a poor fit to the observed
SED. The best-fit model has an age of 1~Gyr, a star formation rate (SFR) of 
$\sim50~\Msunyr$ (assuming a constant star formation history), and a stellar
mass of $3.5\times10^{10}~\Msun$, and no dust. The statistical errors on these
quantities are at the $\sim10\mbox{--}30$\% level; however, the uncertainties 
are dominated by systematics in the foreground removal and lens magnification 
estimate.
The results are similar whether
the Sbc or Scd template is adopted for the foreground.

We next consider a two-component model. This additional flexibility allows for a 
young population to account for the blue UV SED, while an older population 
-- implying significant past star formation -- accounts for the bright IRAC 
fluxes. Given the small number of photometric data points and the
systematic uncertainties mentioned above, the results from 
these fits must be approached with caution. Nonetheless, we find that the best-fit
model includes a young (starburst) component with a fixed age of 5~Myr, a (constant) 
SFR of $\sim100~\Msunyr$, and a stellar mass of $\sim6\times10^8~\Msun$. The evolved 
component has an age of 1~Gyr, SFR~$\sim400~\Msunyr$, and stellar mass 
$\sim3\times10^{11}~\Msun$, again assuming a constant star formation history. The 
evolved population also has $E(B-V)=0.44$.
These values change by only $\sim10$\% whether the Sbc or Scd 
foreground model is used. As can be seen in Figure~\ref{fig:sed}(b), the 
composite model performs much better at reproducing both the blue UV slope and 
the bright IRAC fluxes. For the Sbc template, the single population model has
a total $\chi^2 = 8.2$ for 5 data points, while the composite model has
$\chi^2 = 2.9$. In both cases the number of model parameters is large compared to the
number of data points and thus the $\chi^2$ statistic should be approached with 
caution, but this does indicate a significant improvement with the composite model.

In performing these SED fits on the observed fluxes we are ignoring any 
magnification due to gravitational lensing. As an example, an arc 
redshift of $z=2.5$ would reduce the inferred stellar masses and ages by a 
factor of $\sim8$, suggesting that \lae\ is a moderately massive galaxy with 
a substantial population of older stars. However, an arc redshift of $z\la2$ 
would suggest that the true stellar mass is lower by more than an order of 
magnitude. In our fits we have constrained the stellar age to be less than the
age of the universe at $z=5.4$, but the observed fluxes push the ages close
to this limit. This tension is significantly reduced after correcting for
a factor $>5$ lens magnification which would significantly decrease the
inferred age. Regardless of the total stellar mass,
the strong emission lines and blue UV slope point toward an 
ongoing starburst. \lae\ will be a prime target for JWST, which can provide 
resolved photometry and spectroscopy at rest-frame optical wavelengths --- 
including the Balmer series emission lines --- and greatly improved constraints 
on the current and past star formation in this unusual galaxy.

\subsection{Star formation}

There are multiple indicators available to estimate the star formation rate
in \lae. The UV continuum traces the (relatively) unobscured star formation.
Adopting the continuum fit from the HST imaging ($M_{1350} = -23.0$, 
$\beta_{\lambda} = -2.6$) and applying the relation from \citet{Madau+98}
for a Salpeter IMF and ignoring dust extinction, we obtain a value of 
$\approx 80\Msunyr$. This is consistent with results for the younger 
stellar population in the two-component SED fits presented in 
section~\ref{sec:sedfit}, which are $\sim100\Msunyr$. On the other hand, the 
evolved population from the SED fits has a substantial component of 
dust-obscured star formation, with SFR$~\approx400\Msunyr$ and 
$E(B-V) \approx 0.4$. The luminosity of the Ly$\alpha$ line is also consistent 
with a large SFR. Ignoring dust extinction and applying the \citet{Kennicutt98} calibration for H$\alpha$ (assuming Case B recombination and ignoring lensing) 
to the line luminosity obtained from spectral fitting yields 
SFR(Ly$\alpha$)$~\ga~400\Msunyr$. The line luminosities from both the
narrowband imaging and the model fitting are $\sim$20\% larger; note the 
latter implicitly includes a dust correction.
From both the SED and Ly$\alpha$
profile fitting we conclude that a substantial amount of obscuring dust may
be present, implying a much larger Ly$\alpha$ star formation rate after dust 
correction, which we do not perform here. The gravitational lensing correction 
acts in the opposite direction, and would bring the estimates down by anywhere 
from a factor of 5--25. We conclude that the intrinsic SFR is roughly consistent 
with being in the range $\sim10\mbox{--}100~\Msunyr$.

\section{Conclusions}\label{sec:conclude}

We present extensive observations of an unusual high-redshift lensed galaxy. 
\fullname\ was initially targeted as a quasar candidate within the CFHTLS,
but proved to be an exceptionally bright galaxy at $z=5.426$. Our observations
lead to the following picture of this interesting object:

\begin{itemize}
\item The galaxy is unusually bright for a (likely) non-AGN at $z=5.4$ and
 has one of the largest Ly$\alpha$ fluxes reported to date. The UV continuum
 is easily detected and has SNR$\sim3$ in a moderate-resolution LBT/MODS1
 optical spectrum.
\item The strong Ly$\alpha$ emission from this galaxy suggests a powerful
 starburst and sightlines through which Ly$\alpha$ photons can escape.
 The broad red wing of the line extends over $\sim2000~\kms$ as expected
 from an outflowing shell. The peak of the Ly$\alpha$ emission is redshifted
 by $\sim70~\kms$.
\item Typical AGN emission lines such as \ion{N}{V} and \ion{C}{IV} are
 not detected. However, the \ion{N}{IV}] line is detected. Only a few
 examples of strong \ion{N}{IV}] emitters exist in the literature; this
 may be associated with a large population of massive stars keeping the 
 nebular gas in a hot, highly ionized state.
\item Fits to the observed SED prefer a two-component model, with a young 
 population to fit the blue UV slope found with resolved HST photometry, and 
 a massive, evolved population indicated by bright IRAC fluxes.
\item \lae\ lies behind a massive lensing group and has a spatially 
 proximate, faint galaxy in the foreground. This leads to a highly uncertain
 estimate for the total magnification from gravitational lensing, with values
 in the range $5 \la \mu \la 25$.
 \item The unobscured star formation rate inferred from the Ly$\alpha$ 
 flux, UV continuum, and UV/optical SED of this galaxy all suggest that it 
 is forming stars at a moderate to prodigious rate ($10$--$100~\Msunyr$). 
 However, these  estimates are highly uncertain given the poorly constrained 
 lensing magnification.
\end{itemize}

\lae\ was found serendipitously, but similar galaxies will be readily
discovered in upcoming wide-area surveys such as the HSC-Wide, LSST and WFIRST. 
These bright galaxies are prime targets for detailed spectroscopic studies with
JWST and 30m-class telescopes, from which a more complete picture of the
physical conditions of star-forming galaxies near and within the epoch
of reionization can be formed. Furthermore, some of the more interesting
properties of \lae\ provide a guide for future studies of reionization-era
galaxies. First, the exceptional strength of the Ly$\alpha$ emission combined
with the detection of a high-ionization metal line is suggestive of a 
population of galaxies with observable Ly$\alpha$ emission in the reionization 
epoch due to a hard radiation field that ionizes their local bubble 
\citep[e.g.,][]{Stark+16}. Second, the detection of nebular 
emission lines, in this case \ion{N}{IV}, provides an alternative path to 
obtaining redshifts of galaxies even with the Ly$\alpha$ emission is fully
attenuated by a neutral IGM.

\section{Acknowledgements}

We thank the anonymous referee for a careful read of the manuscript and suggestions that improved its clarity.
This work is based in part on observations made with the NASA/ESA Hubble Space Telescope, obtained at the Space Telescope Science Institute, which is operated by the Association of Universities for Research in Astronomy, Inc., under NASA contract NAS 5-26555. These observations are associated with program GO \#13762
with support provided by NASA through a grant from the Space Telescope Science Institute.
Also based in part on observations made with the Spitzer Space Telescope, which is operated by the Jet Propulsion Laboratory, California Institute of Technology under a contract with NASA. Support for this work was provided by NASA through an award issued by JPL/Caltech (GO 90195).
JPK acknowledges support from the ERC advanced grant LIDA.
ML acknowledges CNRS and CNES for its support.
Observations reported here were obtained at the MMT Observatory, a joint facility of the University of Arizona and the Smithsonian Institution.
This paper used data obtained with the MODS spectrographs built with
funding from NSF grant AST-9987045 and the NSF Telescope System
Instrumentation Program (TSIP), with additional funds from the Ohio
Board of Regents and the Ohio State University Office of Research.
This paper made use of the modsIDL spectral data reduction pipeline developed in
part with funds provided by NSF Grant AST-1108693.
Based in part on observations obtained with MegaPrime/MegaCam, a joint project of CFHT and CEA/IRFU, at the Canada-France-Hawaii Telescope (CFHT) which is operated by the National Research Council (NRC) of Canada, the Institut National des Science de l'Univers of the Centre National de la Recherche Scientifique (CNRS) of France, and the University of Hawaii. This work is based in part on data products produced at Terapix available at the Canadian Astronomy Data Centre as part of the Canada-France-Hawaii Telescope Legacy Survey, a collaborative project of NRC and CNRS. 
This research has made use of the CFHTLS-ZPhots database, operated at CeSAM/LAM, Marseille, France.
This research made use of Astropy, a community-developed core Python package for Astronomy \citep{astropy}.
IRAF is distributed by the National Optical Astronomy Observatory, which is operated by the Association of Universities for Research in Astronomy (AURA) under a cooperative agreement with the National Science Foundation.

{\it Facilities:} 
 \facility{CFHT (Megacam)}, 
 \facility{MMT (Red Channel spectrograph)}, 
 \facility{LBT (MODS1, LUCI1)}, 
 \facility{Spitzer (IRAC)}, 
 \facility{HST (WFPC2,ACS,WFC3)}

\bibliographystyle{mn2e}
\bibliography{z54lae}


\appendix

\section{Additional MODS spectroscopy}\label{sec:appendix}

The MODS1 spectroscopic observations included slits placed on interesting objects
within the field of \lae. These include a long slit aligned with the tangential
arc T1 (masks 505919 and 523405), two slits for each of the radial arc candidates
R1 and R2 (mask 510122), and cluster member galaxies selected through simple colour
cuts. Specifically, the member galaxy candidates were targeted with the criteria
$g-r>1.2$ and $r-i>0.7$ using the CFHTLS photometry. In total these criteria select
662 objects within 6\arcmin\ of the BCG position to a depth of $i<25$. We targeted
52 galaxies within $\sim2.5$\arcmin\ of the BCG, obtaining redshifts for 43 
galaxies with $0.58<z<0.64$ (the BCG redshift is $z=0.616$). A few additional
galaxies that did not meet the colour criteria were also targeted on order to fill
the slitmasks.

\begin{figure}
\centering
\epsscale{1.0}
\plotone{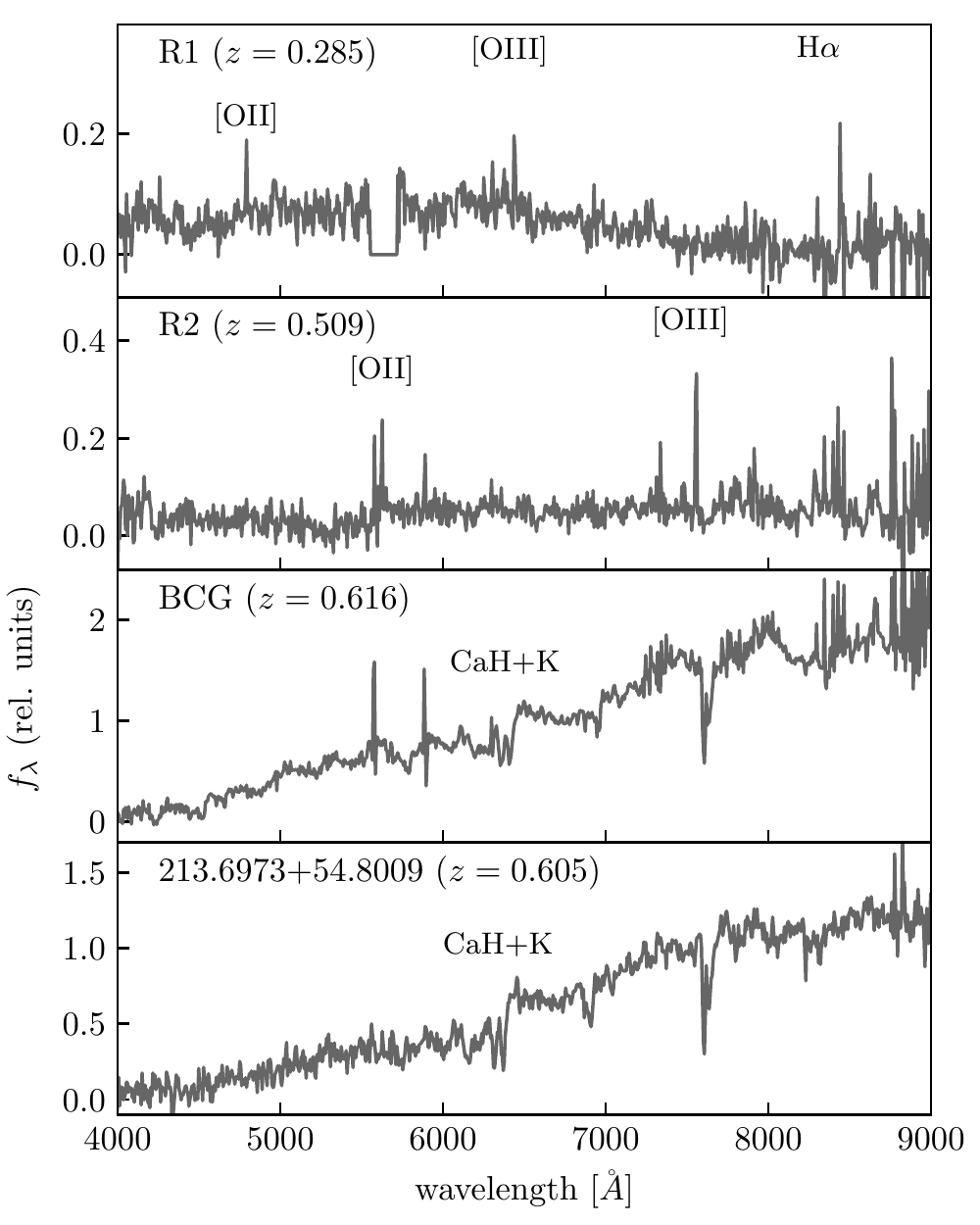}
\caption{Spectra of galaxies in the field of \lae\ targeted on MODS slitmasks.
 Spectra of the two candidate arcs R1 and R2 are presented in the top two
 panels; both are star-forming galaxies with strong emission lines. The
 unambiguous redshifts place them in the foreground of the lensing cluster.
 The third panel from the top shows the BCG spectrum, and the lowest panel
 a randomly selected cluster member galaxy that is representative of the 
 quality of the MODS spectra.
 \label{fig:galspec}
 }
\end{figure}

The full redshift catalog from the MODS1 observations is given in 
Table~\ref{tab:clustertab}. Objects with $z=0$ are confirmed late-type 
stars, while those with an empty redshift entry were not identifiable, usually 
due to low $S/N$. Typical errors on the redshifts are $<20~\kms$. Example MODS 
spectra are presented in Figure~\ref{fig:galspec}.

\begin{figure*}
\centering
\epsscale{1.5}
\plotone{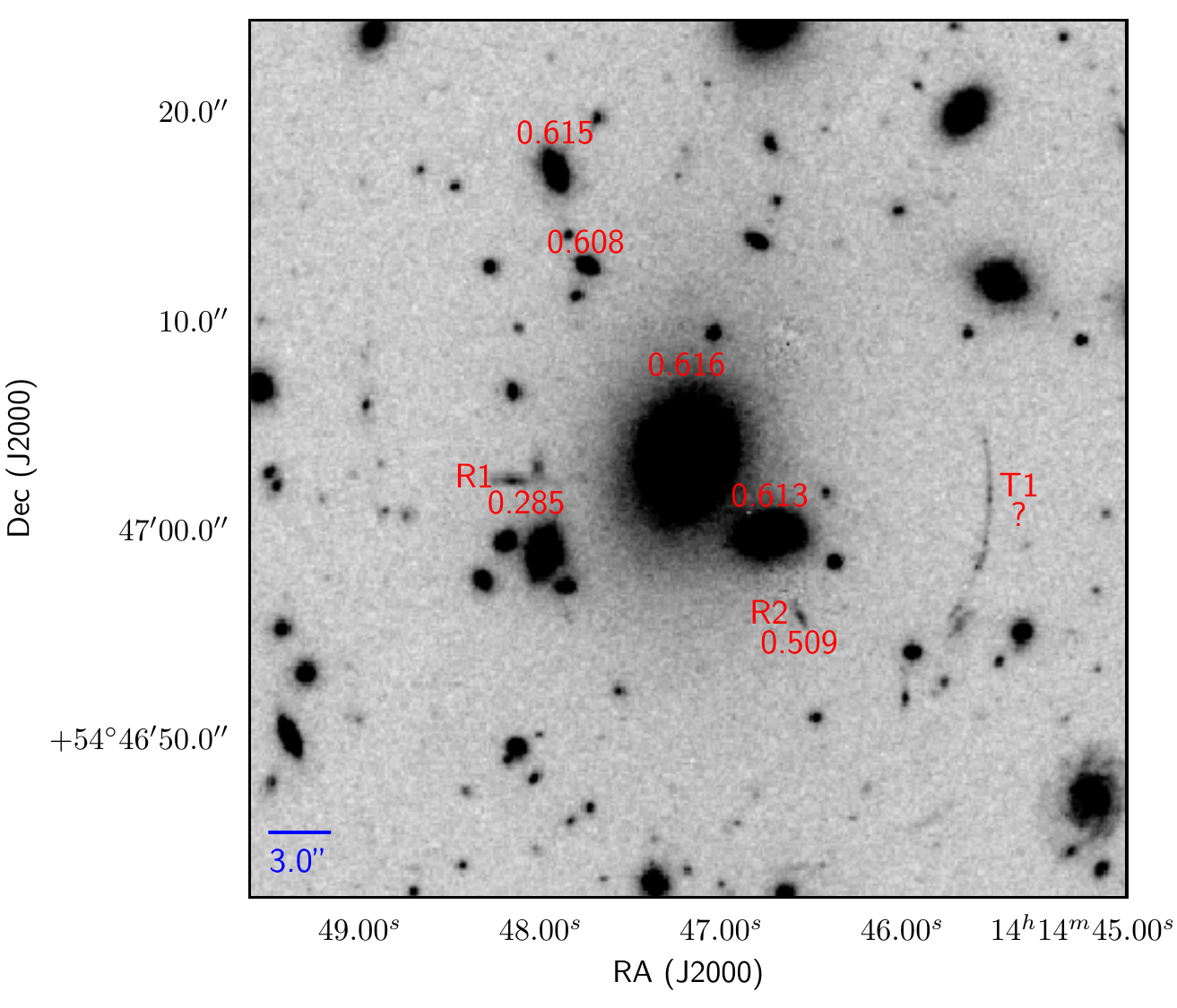}
\caption{Central $\sim$20\arcsec\ of the lensing cluster \slsid\ from the
 HST/WFC3 F125W image. Objects with MODS spectroscopy are labeled with their
 redshift just above the source, except for the candidate lensing arcs which
 are labeled with both their name and the redshift below the name. The data
 are unable to yield a redshift for the tangential arc T1. The two candidate
 radial arcs R1 and R2 are in the foreground of the cluster.
 \label{fig:clusterim}
 }
\end{figure*}

We obtain a redshift of $z=0.285$ for the candidate radial arc R1 from 
\citet{More+12}, based on detections of \ion{O}{III} 5007 and \ion{O}{II} 3727,
and a marginal detection of H$\alpha$. The R1 spectrum may be 
contaminated by a nearby galaxy to the NW (see Fig.~\ref{fig:clusterim})
and as such there remains some possibility that it is at higher redshift.
The candidate arc R2 has a redshift of
$z=0.509$ based on \ion{O}{III} 4959,5007 and \ion{O}{II} 3727. These redshifts
place both of the objects in the foreground of the lensing mass, and thus they
are not lensed arcs. Even with a total exposure time of $\sim9$ hours, we 
were unable to obtain signal on the candidate tangential arc T1. We first 
attempted to collapse the full 14\arcsec\ slit into a single 1D spectrum with
uniform per-pixel weights. Next, we constructed a 1D slit profile matching the 
profile in the HST images in order to assign greater weight to the brighter 
knots associated with the galaxy. Neither method resulted in any detectable 
continuum or line emission. An image of the targets in the cluster centre
is presented in Table~\ref{fig:clusterim}.

\begin{table*}
\caption{Redshifts of field galaxies.}
\label{tab:clustertab}
\begin{tabular}{ccccccccc}
RA (J2000) & Dec (J2000) & $i (AB)$ & $z$ & &
RA (J2000) & Dec (J2000) & $i (AB)$ & $z$ \\
\hline
213.63864 & 54.80231 & 20.37 & 0.61207 & \phantom{aaa} &
213.63982 & 54.80292 & 21.09 & 0.62854 \\
213.64101 & 54.76356 & 21.61 & 0.62484 & \phantom{aaa} &
213.64482 & 54.77097 & 17.18 & 0.00000 \\
213.64774 & 54.80578 & 20.40 & 0.61501 & \phantom{aaa} &
213.64890 & 54.76526 & 20.82 & 0.60856 \\
213.65602 & 54.77777 & 21.88 & 0.59960 & \phantom{aaa} &
213.65744 & 54.81451 & 21.26 & 0.51609 \\
213.66525 & 54.78886 & 19.93 & 0.59645 & \phantom{aaa} &
213.66668 & 54.81920 & 21.41 & 0.61464 \\
213.66811 & 54.80514 & 20.54 & 0.61257 & \phantom{aaa} &
213.66841 & 54.78838 & 19.71 & 0.59668 \\
213.66989 & 54.80754 & 21.24 & 0.85058 & \phantom{aaa} &
213.68693 & 54.79591 & 21.11 & 0.62238 \\
213.68789 & 54.76468 & 20.08 & 0.60376 & \phantom{aaa} &
213.69023 & 54.78886 & 20.95 & 0.61697 \\
213.69038 & 54.80107 & 21.77 & 0.60402 & \phantom{aaa} &
213.69096 & 54.80299 & 21.75 & 0.61283 \\
213.69357 & 54.79611 & 21.00 & 0.61429 & \phantom{aaa} &
213.69402 & 54.78219 & 23.21 & 0.50876 \\
213.69470 & 54.78325 & 19.56 & 0.61347 & \phantom{aaa} &
213.69480 & 54.79008 & 19.83 & 0.62059 \\
213.69486 & 54.75863 & 20.62 & 0.61276 & \phantom{aaa} &
213.69503 & 54.74819 & 21.95 & 0.19803 \\
213.69573 & 54.82074 & 21.76 & 0.61585 & \phantom{aaa} &
213.69662 & 54.78430 & 18.32 & 0.61562 \\
213.69729 & 54.80092 & 20.30 & 0.60472 & \phantom{aaa} &
213.69766 & 54.79988 & 21.78 & 0.61200 \\
213.69860 & 54.80782 & 20.31 & 0.62577 & \phantom{aaa} &
213.69895 & 54.78682 & 21.15 & 0.60836 \\
213.69907 & 54.75195 & 19.29 & 0.00000 & \phantom{aaa} &
213.69965 & 54.78808 & 20.93 & 0.61550 \\
213.69966 & 54.75082 & 21.79 & 0.60437 & \phantom{aaa} &
213.70032 & 54.78400 & 21.64 & 0.28546 \\
213.70149 & 54.79680 & 22.09 & 0.62181 & \phantom{aaa} &
213.70456 & 54.79925 & 22.26 & 0.61868 \\
213.70540 & 54.78141 & 22.18 & 0.61843 & \phantom{aaa} &
213.70575 & 54.78055 & 21.49 & 0.59664 \\
213.70600 & 54.76151 & 19.73 & 0.51899 & \phantom{aaa} &
213.70635 & 54.80830 & 21.62 & 0.61173 \\
213.70646 & 54.78522 & 21.57 & 0.61511 & \phantom{aaa} &
213.70714 & 54.76068 & 21.65 & 0.60930 \\
213.70874 & 54.82911 & 18.55 & 0.00000 & \phantom{aaa} &
213.71126 & 54.74782 & 21.75 & 0.61017 \\
213.71588 & 54.74148 & 18.80 & 0.52287 & \phantom{aaa} &
213.71917 & 54.75171 & 20.90 & 0.56825 \\
213.71998 & 54.75488 & 21.22 & 0.52187 & \phantom{aaa} &
213.72032 & 54.76813 & 21.19 & 0.61002 \\
213.72506 & 54.74987 & 21.16 & 0.61393 & \phantom{aaa} &
213.72995 & 54.77044 & 22.08 & 0.61455 \\
213.73236 & 54.76782 & 21.52 & 0.61254 & \phantom{aaa} &
213.73911 & 54.79662 & 21.19 & 0.61576 \\
213.74338 & 54.79890 & 21.05 & 0.62441 & \phantom{aaa} &
213.74414 & 54.77551 & 19.72 & 0.61317 \\
213.74979 & 54.80938 & 21.63 & 0.61607 & \phantom{aaa} &
\end{tabular}
\end{table*}

\end{document}